\documentclass[twocolumn,
	aps, prd,
	10pt, notitlepage, 
        floats, floatfix,
	amsmath, amssymb, amsfonts, eqsecnum,
	superscriptaddress,
	showpacs, showkeys,
	nofootinbib,
 	longbibliography,
]{revtex4-2}

\usepackage{mathtools}
\usepackage{wrapfig}
\usepackage{graphicx} 
\usepackage[usenames,dvipsnames]{xcolor}
\usepackage{xspace} 
\usepackage{bm} 
\usepackage{amsmath,amsfonts,amssymb,amsthm,mathtools}
\usepackage[utf8]{inputenc} 

\xdefinecolor{mylinkcolor}{rgb}{0,0,0.5}
\usepackage[
	bookmarksnumbered, bookmarksopen, bookmarksopenlevel=2,
	breaklinks=true,
	colorlinks=true, filecolor=mylinkcolor, citecolor=mylinkcolor,
	linkcolor=mylinkcolor, urlcolor=mylinkcolor, menucolor=mylinkcolor,
]{hyperref}

\usepackage{verbatim}
\usepackage{mathrsfs}
\usepackage{color}
\usepackage{enumitem}
\usepackage[dvipsnames]{xcolor}
\usepackage{comment}
\usepackage{tikz}
\usetikzlibrary{decorations.pathmorphing}
\usetikzlibrary{arrows}

\tikzset{snake it/.style={decorate, decoration=snake}}

\usepackage[normalem]{ulem}

\newcommand{\nnm}{\nonumber}

\newcommand{\be}{\begin{equation}}
\newcommand{\ee}{\end{equation}}
\newcommand{\bse}{\begin{subequations}}
\newcommand{\ese}{\end{subequations}}

\newcommand{\mr}{\mathrm}
\newcommand{\tr}{\textrm}
\newcommand{\mc}{\mathcal}

\newcommand{\bpm}{\begin{pmatrix}}
\newcommand{\epm}{\end{pmatrix}}

\newcommand{\AEI}{\affiliation{Max Planck Institute for Gravitational Physics (Albert Einstein Institute), Am M\"uhlenberg 1, Potsdam 14476, Germany}}
\newcommand{\Maryland}{\affiliation{Department of Physics, University of Maryland, College Park, MD 20742, USA}}

\begin{document}

\title{Scattering of gravitational waves off spinning compact objects \\ with an effective worldline theory}
\author{M. V. S. Saketh}
\email{msaketh@aei.mpg.de}
\Maryland\AEI
\author{Justin Vines}
\email{justin.vines@aei.mpg.de}
\AEI
\date{\today}
\begin{abstract}

We study the process, within classical general relativity, in which an incident gravitational plane wave, of weak amplitude and long wavelength, scatters off a massive spinning compact object, such as a black hole or neutron star.  The amplitude of the asymptotic scattered wave, considered here at linear order in Newton's constant $G$ while at higher orders in the object's multipole expansion, is a valuable characterization of the response of the object to external gravitational fields.  This amplitude coincides with a classical ($\hbar\to0$) limit of a quantum 4-point (object and graviton in, object and graviton out) gravitational Compton amplitude, at the tree (linear-in-$G$) level.  Such tree-level Compton amplitudes are key building blocks in generalized-unitary-based approaches to the post-Minkowskian dynamics of binaries of spinning compact objects.  In this paper, we compute the classical amplitude using an effective worldline theory to describe the compact object, determined by an action functional for translational and rotational degrees of freedom, including couplings of spin-induced higher multipole moments to spacetime curvature.  We work here up to the levels of quadratic-in-spin quadrupole and cubic-in-spin octupole couplings, respectively involving Wilson coefficients $C_2$ and $C_3$.  For the special case $C_2=C_3=1$ corresponding to a black hole, we find agreement through cubic-in-spin order between our classical amplitude and previous conjectures arising from considerations of quantum scattering amplitudes.  We also present new results for general $C_2$ and $C_3$, anticipating instructive comparisons with results from effective quantum theories.
\end{abstract}

 
\maketitle

\section{Introduction}

Since Einstein's proposal of general relativity as the correct description of gravity in 1915 \cite{Einstein:1915ca,Einstein:1916vd}, a significant body of research has been developed in understanding and testing its consequences. Even today, more than a century later, there is still work to be done in this regard. In the age of gravitational-wave (GW) astronomy \cite{Abbott:2016blz, TheLIGOScientific:2016pea, TheLIGOScientific:2017qsa, LIGOScientific:2018mvr, Abbott:2020niy}, progress depends on deepening our understanding of the general relativistic two-body problem, as the primary sources of GWs are binary systems of compact objects such as black holes and neutron stars. 
Identifying and characterizing such GW signals requires efficiently generated and highly accurate waveform templates, constructed both from interpolations of large-scale numerical simulations of binary spacetimes and from semi-analytic approximations to binary dynamics.   The demand in accuracy has been further enhanced in recent years, with the promise of upcoming more sensitive detectors \cite{Reitze:2019iox,Punturo:2010zz,Audley:2017drz}, with the exciting possibility of identifying finite size effects such as spin, spin-induced deformations and tidal effects in the future. In this work, we will primarily focus on the inclusion of spin and spin-induced multipole moments into the dynamics of compact bodies in the context of the two-body problem.

Black holes and compact bodies are extended objects, and their surfaces probe very strong gravitational fields. Nevertheless, it can be convenient to treat them as point particles moving along worldlines when describing their dynamics. As long as one is viewing them from afar, a representative worldline can be chosen describing the bulk motion, and the strong gravitation and (possibly complicated) internal structure can be taken into account by letting the ``particle'' couple non-minimally to the external gravitational field. Frameworks which treat the dynamics of extended bodies in this way may be collectively referred to as ``effective worldline theories.'' This framework is especially useful for dealing with spinning bodies, which are necessarily extended. Rotation also tends to deform the body, endowing it with a tower of multipole moments, with the $2^l$-pole moment scaling as $m a^{l}$, where $S= m a$ is the spin (intrinsic) angular momentum and $m$ is the mass of the body (with $c=1$). The problem of including spin and spin-induced multipole moments into the gravitational dynamics of the body has been tackled in various ways historically. Focussing on effective worldline approaches, some of the earliest works are Refs.~\cite{Mathisson:1937zz, Papapetrou:1951pa}, in which the equations of motion for a spinning particle were established at linear order in spin. 

Beyond linear order in spin, one also needs to include the effects of spin-induced multipole moments on the dynamics, starting with a quadrupole moment scaling as $ma^2$. 
One approach is to construct an ansatz for an effective worldline stress-energy tensor for the particle, including spin and higher multipoles, and then appropriate equations of motion follow from the conservation law $\nabla_{\mu}T^{\mu\nu}=0$; see e.g.\ Ref.~\cite{Steinhoff:2009tk,Dixon1964, Dixon1973, Dixon1974}. 
Alternatively, one can construct an ansatz for a worldline action with rotational degrees of freedom and higher multipole couplings. This was approached for a spinning point particle in flat spacetime in Ref.~\cite{Hanson:1974qy}, and generalized for curved space and arbitrary multipolar structure in Ref.~\cite{Bailey:1975fe}. 

In the context of the post-Newtonian (PN) approximation of the two-body problem, the action approach was developed in an effective field theory (EFT) treatment in Ref.~\cite{Porto:2005ac}, which computed the leading order (LO) spin-orbit contributions (at 1.5PN order) and the LO spin$^2$-quadrupole contributions (at 2PN order) to the effective Hamiltonian. The effect of the spin-induced quadrupole moment was included in this work by adding a quadratic-in-spin coupling term, linear in the Riemann curvature tensor, to the action, along with an undetermined coefficient to be fixed by a further matching calculation. The leading contribution of a generic spin-induced quadrupole moment to binary dynamics and gravitational wave form was also computed earlier in Ref.~\cite{Poisson:1997ha}. A significant amount of work 
has been since done in the inclusion of spin into binary dynamics, as reviewed e.g.\ in Refs.~\cite{Blanchet:2013haa,Porto:2016pyg,Levi:2018nxp}, pushing the state-of-the-art for spinning particles to the N$^3$LO spin-orbit \cite{ Levi:2020kvb, Antonelli:2020aeb}, N$^3$LO quadratic-in-spin \cite{Levi:2020uwu, Antonelli:2020ybz, Kim:2021rfj}, NLO cubic-in-spin \cite{Levi:2019kgk}, and NLO quartic-in-spin \cite{Levi:2020lfn}.
Spinning binary dynamics may also be treated in a post-Minkowskian expansion, where one works to all orders in speeds but perturbatively in the strength of the fields, regulated by powers of the gravitational constant $G$, as in the post-Minkowskian worldline approaches in Refs.~\cite{Kalin:2020mvi,Dlapa:2021vgp,Kalin:2020fhe,Mogull:2020sak,Jakobsen:2021smu,Jakobsen:2021lvp,Jakobsen:2021zvh,Jakobsen:2022fcj,Kalin:2022hph,Jakobsen:2022psy}. Although the post-Minkowskian expansion is more suitable for dealing with scattering encounters, one can extract the dynamics of bound binaries as well from the observables associated with scattering encounters. This can be done either via boundary-to-bound maps between scattering and bound observables (see Refs.~\cite{Kalin:2019rwq, Kalin:2019inp, Saketh:2021sri, Cho:2021arx}) or by constructing a Hamiltonian (see Refs.~\cite{Damour:2016gwp,Damour:2017zjx,Vines:2018gqi}).

Apart from worldline based approaches, a significant body of recent research has been dedicated to quantum field theory (QFT) based or amplitude-based methods. In these approaches, the key quantity of interest is usually the 2-to-2 scattering amplitude: two particles (representing massive compact objects) in, and two particles out. Given this amplitude, the relevant scattering observables (see e.g.\ Ref.~\cite{Bjerrum-Bohr:2018xdl,Kosower:2018adc}) or a classical Hamiltonian (see e.g. Ref.~\cite{Cheung:2018wkq}) can be derived from its classical limit. Similar to the worldline approaches, there are multiple sub-classes among the QFT-based methods for tackling the classical two-body problem as well.

 One line of approaches employ an explicit quantum field, with an action that couples it to gravitation. The quanta of this field, in a classical limit, are used as avatars for black holes or other compact bodies. The 2-to-2 scattering amplitude (for two such quanta in, exchanging gravitons, and two quanta out) can be used to derive binary conservative dynamics. For spinless black holes, an effective conservative Hamiltonian describing binary dynamics was derived at 2PM order ($O(G^2)$) in Ref.~\cite{Cheung:2018wkq}, at 3PM in Ref.~\cite{Bern:2019nnu} and at 4PM in Refs.~\cite{Bern:2021dqo, Bern:2021yeh}, where a massive scalar field minimally coupled to gravity was employed. An important early work in the inclusion of spin in these approaches was done in Ref.~\cite{Vaidya:2014kza}, where the scattering amplitude for the scattering of a scalar quantum ($\phi$) and a quantum of a spinning field ($\psi$), $\phi\psi\rightarrow \phi\psi$, was computed and subsequently used to derive a classical Hamiltonain describing their interaction at the leading PN orders. It was seen that the Hamiltonian matched that of a two-body system with one spinning and one spinless black hole up to fourth order in spin when the spinning member was a quantum of a spin-2 field minimally coupled to gravity. The scattering amplitude for the case where both particles were spinning, and subsequently a Hamiltonian describing their dynamics was derived in Ref.~\cite{Bern:2020buy} at linear order in the spin of each particle at 2PM order (or at order $G^2 S$), and to all orders in spin at 1PM order (at $G S^{\infty}$). This was extended in Ref.~\cite{Kosmopoulos:2021zoq} to $G^2 S^2$, and then to $G^2 S^5$ in \cite{Bern:2022kto}.  These works used an arbitrary-spin-$s$ quantum field, coupled to gravity via all symmetry-permitted non-mimimal coupling terms, parametrized by unknown coefficients. These unknown coefficients were fixed for the case of a black hole in Ref.~\cite{Bern:2022kto} by requiring a best-behaved high-energy limit and a certain ``shift symmetry'' conjecture (in addition to matching with known tree-level results).

An important point regarding these approaches (and others discussed below), which is relevant here, is that the 2-to-2 scattering amplitude is constructed from simpler building block amplitudes, namely the (gravitational) Compton amplitude and the 3-point amplitude, via generalized unitarity. The Compton amplitude (here) is the scattering amplitude for a graviton scattering off a massive spinning quantum, and the 3-point amplitude is the vertex for two massive spinning particles joining a graviton and is related directly to the classical particle's linearized gravitational field. These serve as building blocks for the 2-to-2 scattering amplitude, which brings us to another class of amplitude-based methods.

Another series of works has sought to directly fix or constrain the two-body scattering amplitudes or their building blocks (Compton and 3-point), without going through an explicit QFT action functional, primarily focusing on the black hole case. In Ref.~\cite{Arkani-Hamed:2017jhn}, a particular 3-point amplitude (dubbed ``minimal coupling'') for an arbitrary-spin-$s$ particle was singled out by its uniquely well-behaved high-energy limit; a certain (remarkably simple) Compton amplitude consistent with this 3-point was also singled out, though exhibiting spurious poles for spins $s>2$ (eventually corresponding to beyond fourth order in classical spin). These 3-point and Compton amplitudes were  conjectured to correspond to black holes and used to compute the 2-to-2 scattering amplitudes to $G^1S^\infty$ and $G^2S^4$ orders in Ref.~\cite{Guevara:2017csg}.  The 2-to-2 amplitudes were later used to compute a classical aligned-spin scattering angle function in Ref.~\cite{Guevara:2018wpp}, and contributions to binary effective potentials in Ref.~\cite{Chung:2018kqs}, and finally a complete binary Hamiltonian for generic spin orientations in Ref.~\cite{Chen:2021qkk}, all through the same  $G^1S^\infty$ and $G^2S^4$ orders.  
The same Compton amplitude for spins $s\le 2$ from Ref.~\cite{Arkani-Hamed:2017jhn} was re-derived from BCFW recursion applied to the 3-point in Ref.~\cite{Aoude:2020onz} employing a heavy-particle EFT formalism; its double-copy properties and its further applications were studied in Refs.~\cite{Bautista:2019evw, Bautista:2019tdr}, and in Ref.~\cite{Johansson:2019dnu} which first produced the Compton for both helicity configurations from a double copy prescription.

Pushing beyond fourth order in spin, Ref.~\cite{Aoude:2022trd} gave a parametrization of a spurious-pole-free Compton amplitude, using the heavy-particle EFT formalism, while enforcing consistency with general principles and with the ``minimal-coupling'' (black-hole) 3-point ampitudes, \emph{and} while conjecturally imposing a certain ``black-hole spin structure'' observed at lower orders (which for 2-to-2 amplitudes is equivalent to the ``shift symmetry'' posited in Ref.~\cite{Bern:2022kto}).  Constructing the 2-to-2 amplitude from their parametrized Compton, they found that all freedom therein was fixed by additionally requiring a best-behaved high-energy limit (also just as in Ref.~\cite{Bern:2022kto}), producing results at fifth order in spin in complete agreement with those from the simultaneous Ref.~\cite{Bern:2022kto}, as well as results at higher orders in spin.\footnote{The quintic-in-spin Compton as constrained by requiring the best-behaved high-energy limit in Ref.~\cite{Aoude:2022trd} was shown there to be at odds with a Compton derived in Ref.~\cite{Chiodaroli:2021eug} from a Lagrangian for a massive spin-5/2 particle which is uniquely determined by its 3-point satisfying ``the current constraint'' (known from higher-spin theory and necessary for the existence of an underlying unitary theory) and requiring a minimal number of derivatives in the coupling. An interesting question is whether such tension may be alleviated with the generalization of such higher-spin Lagrangians to the large-spin limit.  

It is also of note that the extrapolated ``shift symmetry'' \cite{Bern:2022kto} used (at the level of the 2-to-2 amplitude) to constrain higher-spin tree-level Compton amplitudes has been observed not to hold in general (directly) in the 1-loop-order Compton amplitude as computed recently in Ref.~\cite{Chen:2022yxw}.}  It was argued that these considerations in fact fix the 2-to-2 amplitude at $O(G^2)$ to all orders in spin (while leaving freedom in the Compton to which the 2-to-2 is insensitive), at least for the case of the 2-to-2 amplitude for a spinless particle meeting a spinning black hole, at order $G^2S^\infty$, as explicitly constructed in Ref.~\cite{Aoude:2022thd}.
 
These works represent impressive progress in constraining effective black-hole--graviton amplitudes at higher orders in spin, but it is important to stress their conjectural nature.  While the 3-point amplitudes have been very concretely matched (in many of the above references) with the linearized Kerr metric and the corresponding linearized effective worldline action \cite{Levi:2015msa,Vines:2016qwa,Harte:2016vwo,Vines:2017hyw}, constraints on the black-hole Compton amplitude at higher orders in spin (beyond consistency with the 3-point) have thus far been based on good-amplitude considerations without a direct connection to black hole physics.\footnote{A first confrontation of the black-hole Compton amplitude conjectures with actual black hole physics was carried out in Ref.~\cite{Siemonsen:2019dsu}, which compared the (2-to-2) aligned-spin scattering angle through $G^2S^4$ order from Ref.~\cite{Guevara:2018wpp} with ``self-force'' calculations of the linearized perturbations of a Kerr black hole produced by a small orbiting body.  This provided partial verification of the conjectures but left open freedom which may be best fixed by direct comparison of (classical) Compton amplitudes.}  As suggested in Ref.~\cite{Bautista:2021wfy}, a promising approach is to identify the classical limit of the quantum Compton amplitude with the amplitude for classical scattering of a gravitational plane wave off a Kerr background, determined by solutions of the Teukolsky equation describing linear gravitational perturbations of a spinning black hole.  Such calculations have a long history \cite{Matzner:1977dn,Dolan:2007ut}, but have so far produced explicitly results of the necessary kind only through linear order in spin \cite{Dolan:2008kf}. (The toy model of a massless scalar field scattering off Kerr was studied at higher orders in spin in Ref.~\cite{Bautista:2021wfy}, leaving the gravitational-wave case for future work.)  In the present work, we show that the same (tree-level) ``classical Compton amplitude'' can be usefully computed from an effective worldline action description, not only for black holes, but also for bodies with general spin-induced mutlipole moments.

The universally accessible nature of Compton amplitudes (specifically its classical limit) suggests that it can be a valuable tool in comparison and calibration of different effective approaches (worldline or QFT) and also with the real system of a (say) a Kerr black hole. For instance, both the worldline and the QFT-based methods are essentially effective approaches to model the dynamics of spinning compact bodies, and both approaches have free coefficients that need to be fixed through matching to produce the dynamics of the desired compact body. As described earlier, in the worldline based approaches, this freedom shows up as unspecified multipole moments, or in the worldline stress energy tensor, or as undetermined couplings in the action. In the amplitude-(or QFT-)based methods, this shows up in the freedom to choose the quantum-field and/or in its coupling to gravitation, or in the parametrization of the building block Compton or 3-point amplitudes. However, the relation between the freedom in these two (worldline vs. amplitude) classes of approaches is not easy to relate. It is known (since Ref.~\cite{Bern:2022kto}) for example that there seem to be more free parameters in the action in QFT based approaches compared to worldline-based approaches to the same order in spin. It is thus of great interest to compare common quantities in these two approaches with each other for understanding the effect and physical meaning of these parameters. This is typically done by computation of the scattering angle, or the Hamiltonain, or other quantities relevant to the two-body system. These are however complicated quantities associated with the interaction between two bodies, when the undetermined coefficients themselves parametrize the interaction of just one body with the external gravitational field. Thus, a convenient quantity requiring just one body, and well-defined for both effective approaches and also easily calculable from the full theory of general relativity is desired. The Compton amplitude satisfies all of these requirements.

Finally, the universal accessibility of the Compton amplitude also means that it may be computed via non QFT-based methods and then employed in a QFT-based framework for the computation of (the classical limit of) the two particle scattering amplitude and subsequently dynamics thus enabling the different ways of studying compact bodies to complement and support each other. This is particularly useful if it is sometimes simpler to compute the Compton amplitude in certain approaches over others. Additionally, it serves as a crucial point of comparison to verify the conjectures that are employed in QFT/amplitude-based methods in constructing the Compton amplitude, which is necessary to trust that the resultant dynamics indeed describe the desired compact body.

 In this work, with these motivations in mind, we derive the Compton amplitude for gravitational waves scattering off a stationary-axisymmetric, parity-preserving spinning compact body to $S^3$ for generic values for the undetermined coefficients ($C_2,~C_3$ up to $S^3$, see Sec.~\ref{exec}) in an effective worldline approach. We do this with a multitude of objectives in mind. Primarily, having the Compton amplitude for generic spinning bodies may be subsequently used for deriving the dynamics of generic bodies in the future. Furthermore, this can be quite useful for comparison with the Compton amplitude derived from effective quantum approaches and understanding the relation between the free coefficients in either approach such as understanding the reason behind having more free parameters in the QFT-based approach in Ref.~\cite{Bern:2022kto}. Additionally, when the coefficients $C_2$ and $C_3$ are fixed for the case of a black hole, we obtain a Compton amplitude consistent with that derived in Refs.~\cite{Arkani-Hamed:2017jhn,Aoude:2020onz} thus corroborating (up to $S^3$) the conjecture relied upon in Refs.~\cite{Guevara:2017csg, Guevara:2018wpp, Chung:2018kqs, Chen:2021qkk} that this Compton amplitude properly describes a black hole (in the classical limit). Finally, we aim to propose the Compton amplitude as a valuable tool for comparison and calibration of the diverse approaches currently being used for studying the motion of compact bodies under gravitational interaction.

\paragraph*{Worldline EFT approach to Compton amplitude:}
For the purpose of deriving the classical Compton amplitudes in this work, we treat the compact body as a point particle equipped with spin-induced quadrupole ($\propto S^2$) and octupole ($\propto S^3$) moments in a worldline formalism as laid out in Ref.~\cite{Marsat:2014xea}. We subject this particle to an external linearized gravitational plane wave perturbation with a large wavelength i.e., $\lambda\gg r_{\mr{CB}}$, where $r_{\mr{CB}}$ is of the order of the size of the compact body.
The incident plane wave perturbs the particle's momentum, spin and subsequently the spin-induced multipole moments which in turn perturb the stress-energy tensor of the particle. When the perturbed stress energy tensor is plugged into the Einstein equation, one finds that it generates an additional wave like perturbation in the metric, which forms a part of the scattering of the incident wave. Additionally, owing to the non-linear nature of gravity, the incident wave also scatters of the static gravitational field of the particle which adds another contribution to the scattered wave also through the Einstein equation.
Comparing the total scattered wave to the incident wave gives us the scattering amplitude at linear order in $G$ and up to third order in spin. We find that this is consistent with the Compton amplitude derived in Ref.~\cite{Arkani-Hamed:2017jhn} for the case of black holes. The following section \ref{exec} succinctly summarizes the our results.

\subsection{Summary of framework and results}
\label{exec}
\begin{figure}[h]
\label{scatter}
\includegraphics[width=8 cm]{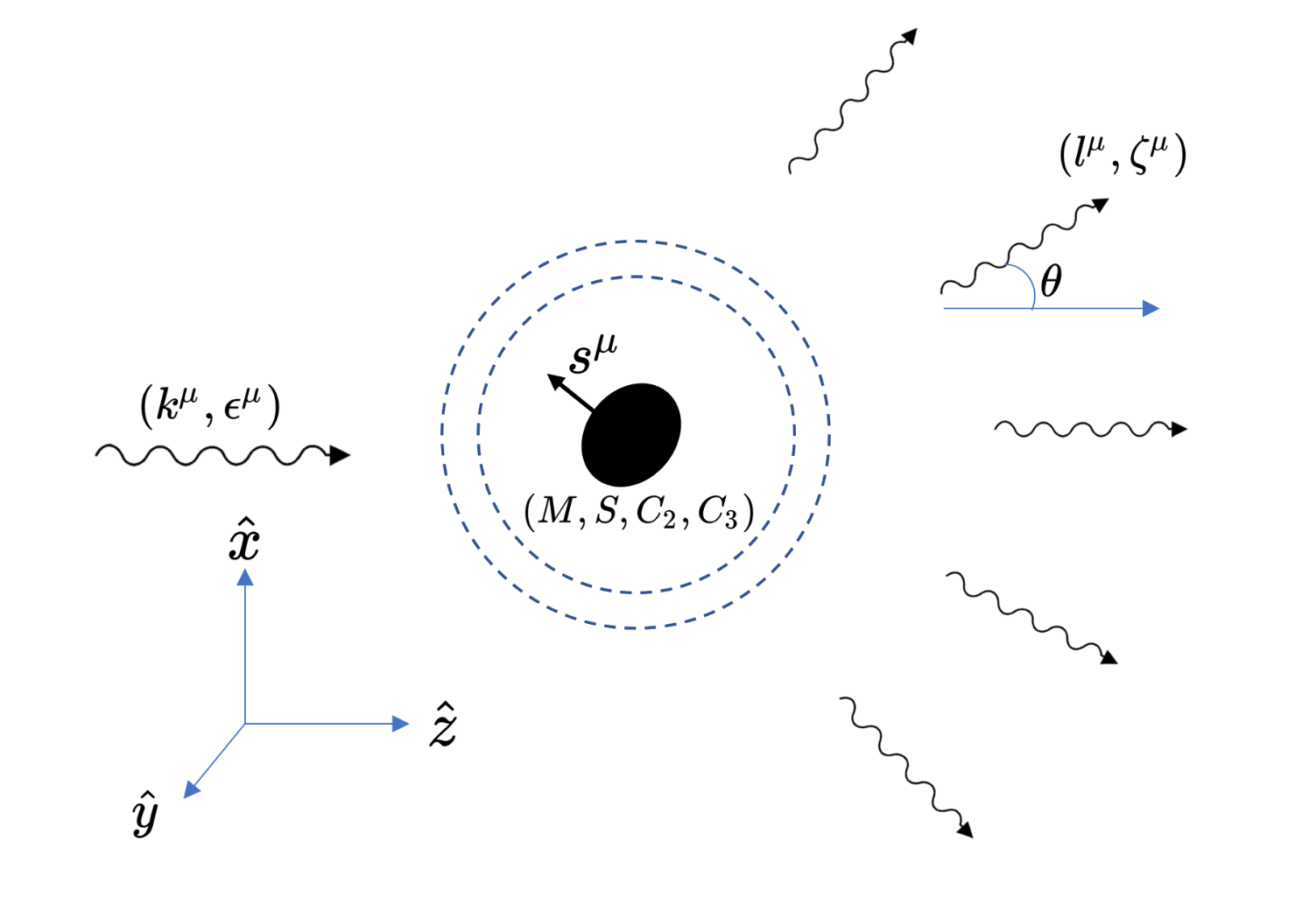}
\caption{The spinning body is characterized by its mass $M$, spin $S$ and spin-induced quadrupole and octupole moments whose magnitudes are controlled by two dimensionless coefficients $C_2$ and $C_3$. An incident wave moving along the $+z$-direction, with wave vector $k^\mu$ and complex polarization vector $\epsilon^{\mu}$, scatters off the particle producing a scattered wave. Far away from the particle, the scattered wave is a superposition of plane waves with wave vector $l^{\mu}$ and polarization vector $\zeta^{\mu}$. We fix the outgoing wave vector to be in the $x$-$z$-plane for convenience. $\theta$ is the angle between the outgoing and incident wave vectors in the rest frame of the particle. }
\centering
\end{figure}

A spinning classical particle with spin-induced quadrupole and octupole moments rests in flat spacetime. It is characterized by its mass $M$, spin $S$, and two coefficients $C_2$, $C_3$. The direction and magnitude of the spin is usually written together in the spin tensor $S^{\mu\nu}$, with $S^{\mu\nu}p_{\nu}=0$ and $S^2 = (1/2)S^{\mu\nu}S_{\mu\nu}$, or the Pauli-Lubanski spin vector $s^{\mu}=-(1/2m)\epsilon^{\mu}{}_{\nu\rho\sigma}p^{\nu}S^{\rho\sigma}$, where $p^{\mu}$ is the momentum of the particle. $C_2$ and $C_3$ control the magnitude of its spin induced quadrupole and octupole moments. In flat space-time, the particle sources a stationary metric perturbation $h^{\mu\nu} = \sqrt{-g} g^{\mu\nu} - \eta^{\mu\nu}$, given in the linearized limit by
\begin{alignat}{3}
h^{\mu\nu}&=-u^{\mu}u^{\nu}[1-\frac{C_2}{2}(a\cdot \partial)^2]\frac{4GM}{r}\nonumber\\&-u^{(\mu}\epsilon^{\nu)}{}_{\rho\alpha\beta}u^{\rho}a^{\alpha}\partial^{\beta}[1-\frac{C_3}{3!}(a\cdot \partial)^2]\frac{4GM}{r},
\label{toy}
\end{alignat}
where $u^{\mu}= (1/m) p^{\mu}$ is the 4-velocity of the particle in flat-space-time and $a^{\mu} = (1/m) s^{\mu}$, with $a\cdot u=0$. The actual non-linear metric sourced by the body approaches this form asymptotically. In the rest frame of the particle, this can be rewritten as
\begin{alignat}{3}
h^{00}&=[1-\frac{C_2}{2}(\vec{a}\cdot \vec{\partial})^2]\bigg(-\frac{4GM}{r}\bigg),~h^{ij} = 0,\nonumber \\& \quad h^{0i} = \epsilon^{i}_{~jk}a^{j}\partial^{k}[1-\frac{C_3}{3!}(\vec{a}\cdot\vec{\partial})^2]\frac{2GM}{r},
\label{toyrest}
\end{alignat}
where $\vec{a}$ can be identified as the spin vector. Comparison of the above expression with the most general asymptotic past-stationary spacetime  given in Eq.(36) of Ref.~\cite{Blanchet:2013haa} shows that the metric perturbation in Eq.~(\ref{toyrest}) can be recovered by imposing stationarity (all multipole moments being constant in time) and identifying the mass-type multipole moments as $I=M,~I_{i}=0,~I_{ij} =- C_2 M a_{\langle i} a_{j\rangle}$, $~I_{ijk}=0$ and current-type multipole moments as $J_{i}=M a_{i},~J_{ij}=0,~J_{ijk}=-(2/3)C_3 M a_{\langle i}a_{j}a_{k\rangle}$ and require that all other multipole moments either vanish or are higher orders in spin.

 If the particle respects parity and only has spin induced multipole moments, the vanishing or irrelevance of all other multipole moments follows trivially from parity invariance. Thus, $M$, $S$, $C_2$ and $C_3$ parametrize the most general axisymmetric parity-preserving stationary linearized metric perturbation. Also note, that adding trace-terms to the multipole moments (e.g., $\delta_{ij} |a|^2$ to $I_{ij}$) only modifies it by terms proportional to $\partial^2 (1/r) \propto \delta^{(3)}(x)$ thus modifying the field only at the location of the particle. Such corrections are expected to be irrelevant since the field at the location of the particle is ill-defined anyway.

We now subject this parity-preserving stationary axisymmetric body to an external linearized gravitational wave perturbation is incident upon it, which is  characterized as\footnote{Starting below, we use the notation $\equiv$ to mean the form of the tensor object in the zero-momentum frame in the coordinate system shown in Fig.~\ref{scatter}. Thus $k^{\mu}\equiv \omega(1,0,0,1)$ means the 4-vector $k$ is pointing along the +ve 'z' direction in this coordinate system.}
\begin{alignat}{3}
& h_{\mr{wave}}^{\mu\nu} = \sqrt{-g}g^{\mu\nu} - \eta^{\mu\nu} = \epsilon\, \varepsilon^{\mu}\varepsilon^{\nu}\exp[i k\cdot x],\nonumber \\&~ k^{\mu} \equiv \omega(1,0,0,1),~\varepsilon^{\mu}\equiv\frac{1}{\sqrt{2}}(0,1,\pm i, 0),\label{hke}
\end{alignat}
where $\epsilon$ is the amplitude of the incoming wave, $k^{\mu}$ is the wave-vector, fixing the direction of propagation, $\omega$ its frequency and $\varepsilon^{\mu}$ is a complex null vector which fixes the helicity of the incoming wave. The metric perturbation due to the incident wave affects the momentum, spin and subsequently the spin-induced multipole moments. These three quantities in turn perturb the stress energy tensor of the particle, leading to a radiated scattered wave through the Einstein equation. Additionally owing to the non-linear nature of gravity, the incident wave also scatters off the static metric perturbation sourced by the particle. Together, these two contributions lead to a scattered wave. The form of the scattered wave, encodes the scattering amplitude. In the particle's rest frame, the scattered wave takes the form (at large distances from the particle)
\begin{alignat}{3}
G~\epsilon~ & h_{\mr{scatter}}^{\mu\nu}(r,\hat{n}) \equiv \epsilon[\mathcal{M}_{\pm  +}(\hat{n}) \zeta^{\mu}_{+2}(\hat{n})\zeta^{\nu}_{+2}(\hat{n}) \\&+ \mathcal{M}_{\pm  -}(\hat{n})(\hat{n}) \zeta^{\mu}_{-2}(\hat{n}) \zeta^{\nu}_{-2}(\hat{n}) ]\frac{\exp[i\omega(r-t)]}{\omega r},\nonumber
\end{alignat}
where $\mathcal{M}_{\pm  \pm}(\hat{n})$ are the scattering amplitudes for a given pair of incoming and outgoing helicities. $\hat{n}$ is the spatial unit vector in a given direction in the rest frame of the particle, $\zeta^{\mu}_{\pm}(\hat{n})$ are the complex null polarization vectors for a wave vector given by $l^{\mu} \equiv \omega(1,\hat{n})$. Thus. $\mathcal{M}_{\pm \pm}(\hat{n})$ is the amplitude for an incident wave with wave vector $k^{\mu}\equiv \omega(1,0,0,1)$, helicity $\pm 2$, scattering into another wave vector $l^{\mu}\equiv \omega(1,\hat{n})\equiv \omega(1,\sin(\theta),\cos(\theta),0)$ with helicity $\pm 2$.

At zeroth order in spin (or for spinless case), We obtain the amplitudes
\begin{alignat}{3}
\mathcal{M}_{ab} &= 	\frac{4 G M \omega^3 (\varepsilon_a \cdot \zeta_b)^2}{(l-k)^2}, \\
\mathcal{M}_{++} &= \mathcal{M}_{--} = G M \omega \frac{\cos^4\big(\frac{\theta}{2}\big)}{ \sin^2\big(\frac{\theta}{2}\big)},
\\ \mathcal{M}_{+-} &= \mathcal{M}_{-+} =G M \omega \sin^2\bigg(\frac{\theta}{2}\bigg),
\end{alignat}
where $\theta$ is the angle between the incoming and outgoing wave vectors ($k^{\mu}$ and $l^{\mu}$ with $2 \omega^2 \sin^2\big(\frac{\theta}{2}\big)=-k\cdot l$).
Note that there is mixing of helicities (non zero $\mathcal{M}_{+-}$ and $\mathcal{M}_{-+}$) even at zeroth order in spin. This is not true for electromagnetic waves scattering off black holes. The above amplitudes are consistent with those obtained in Ref.~\cite{Peters:1976jx} for spinless black holes. Next, to first order in spin (or for a spinning rigid particle with no spin-induced deformation). We obtain the amplitudes 
\begin{alignat}{3}
\label{amp2221s}
\mathcal{M}_{++} =& G M \omega \frac{\cos^4\big(\frac{\theta}
{2}\big)}{\sin^2\big(\frac{\theta}{2}\big)}\Big[\nonumber 1- \tan^2\bigg(\frac{\theta}{2}\bigg)\frac{s^{\mu}}{M}(k_{\mu}+l_{\mu}) \nonumber \\& \quad -i\frac{S_{\mu\nu}k^{\mu}l^{\nu}}{M \omega \cos^2\big(\frac{\theta}{2})}\Big], \\
\label{amp22m21s}
\mathcal{M}_{+-} = & G M \omega \sin^2\bigg(\frac{\theta}{2}\bigg)\bigg[1+\frac{s^{\mu}(k_{\mu}-l_{\mu})}{M}\bigg],
\\ 
\label{ampm22m21s} \mathcal{M}_{--} =& \bar{\mathcal{M}}_{++}(S^{\mu\nu}\rightarrow - S^{\mu\nu}, ~s^{\mu}\rightarrow -s^{\mu}), \\
\label{ampm2221s}\mathcal{M}_{-+} =& \bar{\mathcal{M}}_{+-}(S^{\mu\nu}\rightarrow  -S^{\mu\nu}, ~s^{\mu}\rightarrow -s^{\mu}),
\end{alignat}
where the spin tensor $S^{\mu\nu},~S^{\mu\nu}p_{\nu}=0$, and the Pauli-Lubanski spin vector $s^{\mu}$, appear for the first time.  A bar denotes complex conjugation.  Note that the spin tensor/vector appearing in the amplitude is that of the initial undisturbed particle, prior to the wave like perturbation. We find that the amplitudes are related to their helicity-flipped versions, via a spin-flip and complex conjugation. This is because the helicities of the incoming and outgoing gravitational wave can be flipped via a combination of parity and time-reversal operation on the whole system, which flips the spin $(s^{\mu},~ S^{\mu\nu})$, complex conjugates the amplitudes $(\mathcal{M}\rightarrow \bar{\mathcal{M}})$ but leaves the momenta ($k^{\mu},~l^{\mu}$) unchanged. The differential cross section resulting from these amplitudes agrees with Eq.~(3) of \cite{Dolan:2008kf}, derived there from black hole perturbation theory.

The coefficients $C_2$ and $C_3$ associated with the spin-induced quadrupole ($\propto S^2$) and octupole ($\propto S^3$) moments finally show themselves at second and third order in spin respectively. For the case of a black hole, $C_2=C_3=1$, the scattering amplitudes to third order
are found to be simply the exponentiated version of the first order in spin amplitudes (in Eqs.~(\ref{amp2221s}-\ref{ampm2221s})). We get (to third order in spin)
\begin{alignat}{3}
\label{amp2223s}
\mathcal{M}_{++} &= G M \omega \frac{\cos^4\big(\frac{\theta}{2}\big)}{\sin^2\big(\frac{\theta}{2}\big)} ~ \exp\bigg[{-}\frac{s^{\mu}}{M}(k_{\mu}+l_{\mu})\tan^2\big(\frac{\theta}{2}\big)\nonumber \\& \quad - \frac{i}{M \omega \cos^2(\frac{\theta
}{2})} S^{\mu\nu}k_{\mu}l_{\nu}\bigg],\\\label{amp22m23s}
\mathcal{M}_{+-} &= G M \omega \sin^2\bigg(\frac{\theta}{2}\bigg)~\exp\bigg[\frac{s^{\mu}}{M}(k_{\mu}-l_{\mu})\bigg],\\ 
\label{ampm22m23s} \mathcal{M}_{--} &= \bar{\mathcal{M}}_{++}(S^{\mu\nu}\rightarrow -S^{\mu\nu}, ~s^{\mu}\rightarrow -s^{\mu}), \\
\label{ampm2223s}\mathcal{M}_{-+} &= \bar{\mathcal{M}}_{+-}(S^{\mu\nu}\rightarrow -S^{\mu\nu}, ~s^{\mu}\rightarrow -s^{\mu}),
\end{alignat}
which matches the amplitude given in spinor helicity formalism in Ref.~\cite{Aoude:2020onz}, obtained from Ref.~\cite{Arkani-Hamed:2017jhn} using QFT-based methods.

 The amplitudes do not exponentiate when $C_2$ and $C_3$ are generic. Thus, they correct the amplitudes in Eqs.~(\ref{amp2223s}), (\ref{amp22m23s}), (\ref{ampm22m23s}), (\ref{ampm2223s}) by adding terms proportional to $(C_{2,3}-1)^{n}$. The general expressions for the amplitudes with generic $C_2$ and $C_3$  for generic polarization vectors are given in the Appendix.  The expressions for the helicity-conserving amplitudes ($\mc M_{++}$ and $\mc M_{--}$) for generic $C_2$ and $C_3$ after substitution of polarization vectors can be simplified considerably with help of the vector
\begin{alignat}{3}
w_{\mr{S}}^{\mu} &= \frac{1}{2 \omega\cos^2\big(\frac{\theta}{2}\big)}[\omega(k^{\mu}+l^{\mu})	- i  \epsilon^{\mu}_{~\alpha\beta\gamma}k^{\alpha}l^{\beta}u^{\gamma}],
\label{ssimp} \nonumber \\&
\end{alignat}
which coincides with half the complex conjugate the vector $w^\mu$ from Ref.~\cite{Aoude:2020onz}.
We also define $a^{\mu}=s^{\mu}/m$. Then, in terms of $w_{\mr{S}}^{\mu}$, $k^{\mu}$, $l^{\mu}$, and $a^{\mu}$, the helicity-conserving amplitudes for the BH case $C_2=C_3=1$ can be rewritten as
\begin{alignat}{3}
\mathcal{M}_{++} &=  GM \frac{\cos^4\big(\frac{\theta}{2}\big)}{\sin^2\big(\frac{\theta}{2}\big)} ~ \exp[a\cdot (k+l-2w_{\mr{S}})], \\
\mathcal{M}_{--} &= \bar{\mathcal{M}}_{++}(a^{\mu}\rightarrow -a^{\mu}).
\end{alignat}

The helicity-conserving scattering amplitudes for a spinning particle, with spin-induced quadrupole moments with generic $C_2$ and $C_3$ to third order in spin, is given by 
\begin{alignat}{3}
\label{keyqn}
& \nonumber \mathcal{M}_{++} =G M \omega\frac{\cos^4(\theta/2)}{\sin^2(\theta/2)}\bigg(\exp[a\cdot ( k +l-2 w_{\mr{S}})]
\\ \nonumber
&+\frac{C_2-1}{2}\{[(k-w_{\mr{S}})\cdot a]^2+[(l-w_{\mr{S}})\cdot a]^2 \}
\\
& \nonumber +\frac{C_2-1}{2}[(k-w_{\mr{S}})\cdot a][(l-w_{\mr{S}})\cdot a][(k+l-2w_{\mr{S}})\cdot a]
\\ \nonumber
&- (C_2-1)^2[(k-w_{\mr{S}})\cdot a][(l-w_{\mr{S}}	)\cdot a](w_{\mr{S}}\cdot a) 
\\
 &  + \frac{C_3-1}{6}\{[(k-w_{\mr{S}})\cdot a]^3+[(l-w_{\mr{S}})\cdot a]^3\}\bigg). \nonumber \\& \\
& \mathcal{M}_{--}  = \bar{\mathcal{M}}_{++}(a^{\mu}\rightarrow -a^{\mu}).
\end{alignat}
The expressions and the corresponding basis for simplification of helicity-reversing amplitudes is given later in Sec.~\ref{result} in Eq.~(\ref{keyqn2}). Having compactly summarized the key results, we now proceed to the main part of the text.

In Sec.~\ref{motion}, the framework describing the dynamics of spinning black holes in a worldline EFT framework is presented, along with the Mathison-Papapetrou-Dixon (MPD) equations of motion at cubic order in spin and the expressions for the spin-induced quadrupole and octupole moments.  In Sec.~\ref{linkerr}, The expression for the skeletonized stress energy tensor of the black hole is presented and the static metric sourced by the unperturbed spinning black hole is derived. Further, it is explained how the comparison of static metric with the linearized Kerr metric (up to a gauge transformation) can be used to fix the unknown coefficients in the spin-induced multipole moments. In Sec.~\ref{perturber}, the properties and representation of the incident plane wave and its amplitude, polarization and frequency are presented. In Sec.~\ref{corrall}, The computation of the perturbation to black hole's spin, velocity, momentum, spin-induced multipole moments and stress-energy tensor due to the incident plane wave is explained. In Sec.~\ref{scwan}, the methodology for solving for the scattered wave is presented, as well as the relation between the scattered wave and the scattering amplitudes. In Sec.~\ref{result}, the expressions for the scattering amplitudes are presented. Finally, we conclude in Sec.~\ref{ftrwork}.

\section{Setup}
\label{main}

\subsection{The worldline action for the particle}
\label{motion}
Let  $z^{\mu}(\tau)$  denote the worldline of the particle, and $dz^{\mu}(\tau)/d\tau=u^{\mu}$ denote the the 4-velocity. Here $\tau$ is the parameter used to characterize the worldline and not constrained to be the proper time in the action. At the level of equations of motion (eom), one can fix $\tau$ to be the proper time by imposing $u^{\mu}u_{\mu}=-1$. To describe the rotation and subsequently the spin of the particle, we attach to the particle a body-fixed tetrad $\epsilon_A{}^{\mu}$ satisfying
\begin{alignat}{3}
\epsilon^{A\mu}(\tau)\epsilon^{B}{}_{\mu}(\tau) = \eta^{AB},~
\epsilon^{A}{}_{\mu}(\tau) \epsilon_{A\nu}(\tau) = g_{\mu \nu}(z(\tau)), \nonumber \\&
\end{alignat}
i.e. orthonormality and completeness respectively. The body fixed tetrad spins along with the particle and thus the variation of the tetrad $\epsilon_{A}{}^{\mu}$ with $\tau$ encodes the particle's rotational motion. To see this more clearly, one can also define a global background tetrad $e_{a}{}^{\mu}(x)$ also satisfying orthonormality ($e^{a\mu}e^{b}{}_{\mu} = \eta^{ab}$) and completeness [$e^{a}{}_{\mu}(x) e_{a\nu}(x) = g_{\mu \nu}(x)$] and relate it to the body-fixed tetrad via a Lorentz transformation,
\begin{alignat}{3}
\epsilon_{A}{}^{\mu} &= \Lambda_{A}{}^{a}e_a{}^{\mu}, \\ \Lambda_{Aa}\Lambda_B{}^a{} = \eta_{AB} \nonumber ,&\quad \Lambda_{Aa}\Lambda^A{}_b{}=\eta_{ab}.  
\end{alignat}
The Lorentz matrices $\Lambda^a{}_A$ contain six degrees of freedom, corresponding to 3 boost and 3 rotational degrees of freedom. However, since we want to use the body fixed tetrad to deal with the rotation of the body, we want to eliminate the boost degrees of freedom. In the non relativistic case, this is trivially accomplished by choosing the boost degrees of freedom such that $\Lambda$ maps the background time vector $e_{0}{}^{\mu}$ to the 4-velocity of the centre-of-mass of the body. However, there is no unique notion of centre of mass in relativistic case and thus there is an ambiguity in choosing a worldline $z^{\mu}(\tau)$ to represent the motion of the compact body. This can only be resolved by imposing a spin-supplementary condition (SSC), which removes three degrees of freedom. We will do that later, at the level of equations of motion. 

For now, we proceed with describing the degrees of freedom that can enter our action. We define the covariant angular velocity as 
\begin{alignat}{3}
\Omega^{\mu \nu} = \epsilon_{A}{}^{\mu} \frac{D \epsilon^{A\nu }}{D\tau}.
\end{alignat}
In the non relativistic case, in the rest frame of the centre-of-mass of the body, this is equal to the 3-D dual of the angular velocity. ($\Omega^{ij} \sim \epsilon^{ijk}w_k$, $i,j=1,2,3$). The dynamics of the particle should be independent of the choice of orientation of the body fixed frame tetrad vectors ($\epsilon_A{}^{\mu}$), thus the body fixed tetrad enters the worldline action only through the angular velocity $\Omega^{\mu\nu}$. For a structureless spinning particle (i.e., one without spin-induced multipole moments), the position $z^{\mu}(\tau)$ only enters the action through the metric $g_{\mu\nu}$. We implement the presence of finite-size effects in the action by allowing  further dependence on the curvature tensor and its covariant derivativies. Thus, our ansatz for the action is 
\begin{alignat}{3}
S[z^{\mu}(\tau),\epsilon_A{}^{\mu}(\tau)] = \int d\tau L[u^{\mu},\Omega^{\mu\nu},g_{\mu \nu}, R_{\mu\nu\rho\sigma},\nabla_{\lambda}R_{\mu\nu\rho\sigma}]. \nonumber \\&
\end{alignat}
This ansatz is sufficient to derive the equations of motion (see Ref.~\cite{Marsat:2014xea}). Here, we simply write them down. We first define the quantities 
\begin{alignat}{3}
\label{defn}
& p_{\mu}  = \frac{\partial L}{\partial u^{\mu}},\quad S_{\mu\nu} = 2 \frac{\partial L}{\partial \Omega^{\mu\nu}},\quad J_\mr{Q}^{\mu\nu\rho\sigma} = -6 \frac{\partial L}{\partial R^{\mu \nu \rho\sigma}},\nonumber\\ & J_\mr{O}^{\lambda\mu\nu \rho\sigma}=-12 \frac{\partial L}{\partial \nabla_{\lambda}R_{\mu\nu\rho\sigma}}.
\end{alignat}
where $p_{\mu}$ and $S_{\mu\nu}$ are identified with the physical 4-momentum and spin tensor of the black hole respectively, related to the black hole's angular momentum as $S^{\mu\nu}S_{\mu\nu}=2 J^2$, where $J$ is the magnitude of the intrinsic angular momentum of the spinning black hole. $J_\mr{Q}^{\mu\nu\rho\sigma}$ and $J_\mr{O}^{\lambda\mu\nu\rho\sigma}$ are the quadrupole and octupole\footnote{Strictly speaking, the quadrupole moment can also couple with the covariant derivative of the Riemann curvature tensor, and thus the octupole moment $J^{\lambda\mu\nu\rho\sigma}_{\mr{O}}$ as defined in Eq.~(\ref{defn}) may not be a pure octupole moment. We will however continue to refer to $J_\mr{Q}^{\mu\nu\rho\sigma}$ and $J_\mr{O}^{\lambda\mu\nu\rho\sigma}$ as quadrupolar and octupolar moments respectively. } moments respectively. In this manner, as mentioned before, allowing the action to depend on $R^{\mu\nu\lambda\rho}$ and its covariant derivatives implement finite size effects. We will neglect all higher multipole moments in this work. The equations of motion are then given by
\begin{alignat}{3}
\label{mots}
\frac{D S^{\mu\nu}}{D \tau}&  - 2 p^{[\mu}u^{\nu]} = N^{\mu\nu} = \frac{4}{3}R^{[\mu}_{~\lambda \rho \sigma}J_{\mr{Q}}^{\nu]\lambda \rho\sigma}\\&\nonumber  + \frac{2}{3}\nabla^{\lambda}R^{[\mu}_{~\tau\rho\sigma}J_{O\lambda}^{~~\nu]\tau \rho\sigma} + \frac{1}{6}\nabla^{[\mu}R_{\lambda \tau \rho \sigma}J_{\mr{O}}^{\nu]\lambda \tau \rho \sigma}, \\
\label{motp}
 \frac{D p_{\mu}}{D \tau} & +\frac{1}{2} R_{\mu \nu \rho \sigma} u^{\nu} S^{\rho \sigma}= F^{\mu} =   - \frac{1}{6} J_{\mr{Q}}^{\lambda \nu \rho \sigma} \nabla_{\mu} R_{\lambda \nu \rho \sigma} \nonumber \\& - \frac{1}{12} J_{\mr{O}}^{\tau \lambda \nu \rho \sigma}\nabla_{\mu} \nabla_{\tau} R_{\lambda \nu \rho \sigma}.
\end{alignat}
Solving these equations in the presence of an incident wave will reveal the perturbation to the spin and motion of the black hole. However note that these equations are not sufficient to solve for $p^{\mu}$ $S^{\mu\nu}$ and $u^{\mu}$. We have 10 equations of motion, but 14 quantities. Thus we need four constraints. One constraint is obtained from the normalization condition $u^{\mu}u_{\mu}=-1$. Three more constraints are obtained from the spin-supplementary condition (SSC), which eliminates the boost degrees of freedom in the Lorentz matrices as mentioned before. In this work, we choose the covariant SSC defined by $S^{\mu\nu}p_{\nu}=0$. This is equivalent to the requirement that the mass-dipole moment of the body vanish in the zero-momentum frame of the body. The SSC helps fix the relation between $p^{\mu}$ and $u^{\mu}$ thus completing the set of equations. More generally, one can work with an action that has an additional Gauge invariance associated with changing the SSC and a corresponding shift in the choice of the worldline (see Ref.~\cite{Steinhoff:2015ksa}). However, it is much more convenient to work directly with Covariant SSC at the level of equations of motion (equivalently via a Lagrange multiplier in the action). The two approaches are physically equivalent (see Ref.~\cite{Vines:2016unv}).

In this work, we only work with spin-induced moments as opposed to tidally-induced moments. Thus, the multipole moments cannot depend on the external curvature or it's derivatives, but only upon the spin of the black hole and the momentum (or 4-velocity). Since we are only interested to third order in spin in this paper, since the multipole moments start at $S^2$, we can conveniently switch $p^{\mu}$ with $m u^{\mu}$ in the expression for multipole moments since the $p$ is aligned with $u$ until $S^2$. The indices of the multipole moments obey the same symmetries as that of the Riemann tensor and its derivatives by definition (see Eq.~(\ref{defn})). Finally, the traces of the quadrupole moment are irrelevant, since all dependence in the action on $R^{\mu\nu}, R$ (traces of the Riemann tensor) can be removed as their dependence can be absorbed into a redefinition of variables (see Refs.~\cite{Goldberger:2004jt,Damour:1998jk}). The same reasoning (along with the Bianchi identity) implies that the traces of octupole moment should be irrelevant as well. Thus, we can modify both $J^{\mu\nu\rho\sigma}$ and $J^{\lambda\mu\nu\rho\sigma}$ upto trace terms to make them simpler.

These considerations along with the covariant SSC constraint ($S^{\mu\nu}p_{\nu}\approx m S^{\mu\nu}u_{\nu} + O(S^3)=0$) were used to fix their form in Ref.~\cite{Marsat:2014xea} as given below, 
\begin{alignat}{3}
\label{quad}
& J_{\mr{Q}}^{\mu\nu\rho\sigma} = \frac{3C_2}{m} u^{[\mu} S^{\nu]\lambda}S_{\lambda}^{~[\rho}u^{\sigma]},\\  & J_{\mr{O}}^{\lambda \mu \nu \rho\sigma} = \frac{C_3}{4 m^2}[\Theta^{\lambda[\mu}u^{\nu]}S^{\rho\sigma} + \Theta^{\lambda[\rho}u^{\sigma]}S^{\mu\nu}-\Theta^{\lambda[\mu}S^{\nu]\rho}u^{\sigma} \nonumber \\& -\Theta^{\lambda[\rho}S^{\sigma][\mu}u^{\nu]}-S^{\lambda[\mu}\Theta^{\nu][\rho}u^{\sigma]}-S^{\lambda[\rho}\Theta^{\sigma][\mu}u^{\nu]}],
\label{octu}
\end{alignat}
where $\Theta^{\mu\nu}= S^{\mu \lambda}S^{\nu}_{~\lambda}$, $m=\sqrt{-p^2}$ and the prefactors have been chosen such that $C_2=C_3=1$ for Kerr black holes (see subsection.~\ref{linkerr}).

As an explicit check that any additions to the above expressions for multipole moment tensors that are either trace terms or terms that violate the SSC are irrelevant for the final result, we added some additional terms to the multipole tensors with arbitrary coefficients and ensured that the results were in line with expectations. For example, we used the following modified expression for the quadrupole moment tensor,
\begin{alignat}{3}
J_{Q}^{\mu\nu\rho\sigma} &= \frac{3C_2}{m} u^{[\mu} S^{\nu]\lambda}S_{\lambda}^{~[\rho}u^{\sigma]} + \alpha S^{\alpha\beta}S_{\alpha\beta}g^{\mu\rho}u^{\nu}u^{\sigma}\nonumber \\& +\beta S^{\mu\alpha}S^{\rho}_{~\alpha}g^{\nu\sigma} + \sigma S^{\alpha\beta}S_{\alpha\beta} g^{\mu\rho}g^{\nu\sigma} + \frac{3 H_2}{4m}S^{\mu\rho}S^{\nu\sigma}, \nonumber \\&
\end{alignat}
where $\alpha,~\beta,~\gamma$  are coefficients next to various non-vanishing traces of the original $C_2$ term. $H_2$ is next to a term that is identical to the original $C_2$ term up to trace-terms and terms violating  the covariant SSC constraint. Thus, the final Compton amplitude should be independent of $\alpha,~\beta,~\sigma$ and should only depend on the combination $C_2+H_2$. This is indeed what we found. We added similarly parametrized corrections to the octupolar moment tensor and observed that the final Compton amplitude was appropriately in line with the expectation that we can neglect trace terms and consistently work with the covariant SSC constraint.

$\phantom{yo}$

\subsection{Stress energy tensor and the static gravitational field}
\label{linkerr}

The particle has a stress energy tensor arising from its mass, spin and the spin-induced multipole moments. The expression for the stress energy tensor can obtained from the action. We first rewrite the action as
\begin{alignat}{3}
S[z^{\mu}(\tau),\epsilon_A{}^{\mu}(\tau)]=\int d^4 x \int d\tau ~L~\delta^{(4)}(x-z(\tau)), \nonumber \\&
\end{alignat}
and then the stress energy tensor is given simply by 
\begin{alignat}{3}
T^{\mu\nu} = \frac{1}{\sqrt{-g}} \frac{\delta S}{\delta g_{\mu\nu}},
\end{alignat}
where all dependency (implicit and explicit) of the action on $g_{\mu\nu}$ needs to be taken into account during the variation. This variation was performed in Ref.~\cite{Marsat:2014xea} to obtain Eqs.~(\ref{stress}).
\onecolumngrid
\begin{subequations}
\begin{alignat}{3}
T^{\mu \nu} & = T^{\mu \nu}_{\mr{pole-dipole}} + T^{\mu \nu}_{\mr{quadrupole}}+T^{\mu\nu}_{\mr{octupole}}, \\ T^{\mu \nu}_{\mr{pole-dipole}} & = \int d\tau p^{(\mu} u^{\nu)} \frac{\delta^{(4)}(x-z)}{\sqrt{-g}} - \nabla_{\rho} \int d\tau S^{\rho (\mu} u^{\nu)}\frac{\delta^{(4)}(x-z)}{\sqrt{-g}}, \\
T^{\mu \nu}_{\mr{quadrupole}} & = \int d\tau \frac{1}{3} R^{(\mu}{}_{\lambda \rho \sigma}J_{\mr{Q}}^{\nu) \lambda \rho \sigma}\frac{\delta^{(4)}(x-z)}{\sqrt{-g}} - \nabla_{\rho} \nabla_{\sigma}\int d\tau \frac{2}{3} J_{\mr{Q}}^{\rho (\mu \nu) \sigma} \frac{\delta^{(4)}(x-z)}{\sqrt{-g}},\\
T^{\mu \nu}_{\mr{octupole}} &= \int d\tau\Bigg[\frac{1}{6} \nabla^{\lambda} R^{(\mu}{}_{\xi \rho \sigma}J_{\mr{O}\lambda}{}^{\nu) \xi \rho \sigma} + \frac{1}{12} \nabla^{(\mu} R_{\xi \tau \rho \sigma}J_{\mr{O}}^{\nu) \xi \tau \rho \sigma}\Bigg]\frac{\delta^{(4)}(x-z)}{\sqrt{-g}} \\ & + \nabla_{\rho}\int d\tau\Bigg[-\frac{1}{6}R^{(\mu}{}_{\xi \lambda \sigma}J_{\mr{O}}^{|\rho|\nu)\xi\lambda\sigma} - \frac{1}{3}R^{(\mu}{}_{\xi \lambda \sigma}J_{\mr{O}}^{\nu)\rho \xi \lambda \sigma} + \frac{1}{3}R^{\rho}{}_{\xi \lambda \sigma}J_{\mr{O}}^{(\mu \nu)\xi \lambda \sigma}\Bigg]\frac{\delta^{(4)}(x-z)}{\sqrt{-g}} \nonumber\\ &+ \nabla_{\lambda}\nabla_{\rho}\nabla_{\sigma} \int d\tau \frac{1}{3} J_{\mr{O}}^{\sigma \rho (\mu \nu) \lambda}\frac{\delta^{(4)}(x-z)}{\sqrt{-g}}.\nonumber
\end{alignat}
\label{stress}
\end{subequations}
\twocolumngrid
Eqs.~(\ref{stress}) contains the stress energy tensor for a spinning particle with spin-induced quadrupole and octupole moments at all times. We want to compute the static metric in harmonic Gauge ($\partial_{\mu}h^{\mu\nu}=0$) sourced by the black hole in flat background space-time. Thus, we disregard the curvature dependent terms in the stress energy tensor and substitute it in the Einstein equation. 
\begin{alignat}{3}
G^{\mu\nu} &= - \frac{8 \pi G}{c^4} T^{\mu\nu}|_{\text{in flat space-time}}, \\
\label{flatstress}
& T^{\mu\nu}= \int d\tau [m u^{\mu} u^{\nu}  - \partial_{\rho} S^{\rho (\mu} u^{\nu)}\\& -\partial_{\rho} \partial_{\sigma} \frac{2}{3} J_{\mr{Q}}^{\rho (\mu \nu) \sigma} + \partial_{\lambda}\partial_{\rho}\partial_{\sigma}  \frac{1}{3} J_{\mr{O}}^{\sigma \rho (\mu \nu) \lambda}]\delta^{(4)}(x-u\tau), \nonumber
\end{alignat}
where we have used $z^{\mu}= u^{\mu} \tau$ which holds as the motion is uniform and $p^{\mu} = m u^{\mu}$, which holds true in flat space even for spinning particles in the covariant SSC. We will only work at linear order in $G$ and thus linearize the Einstein equation by substituting $ G h^{\alpha \beta} = \sqrt{-g} g^{\alpha \beta} - \eta^{\alpha \beta}$ and discarding all $O(G^2)$ terms. This yields the linearized Einstein equation
\begin{equation}
\Box h^{\alpha \beta}_{\mr{static}} = \frac{16 \pi }{c^2}T^{\alpha \beta},
\end{equation}
where $\Box= \eta^{\mu\nu}\partial_{\mu}\partial_{\nu}$.
It is easiest to solve in the rest frame of the unperturbed (by external curvature) black hole since the metric is then static i.e. it does not vary with background time. In this frame, we have $\dot{z}^{\mu} = u^{\mu} \equiv (1,0,0,0)$, $S^{0 \nu}\equiv 0$ \footnote{We are using $\equiv$ symbol to denote the expressions in the black hole's rest frame, as defined by its 4-velocity prior to any external disturbance.} and the equation takes the form
\begin{alignat}{3}
\delta^{ij}\partial_{i}\partial_{j} h^{\mu \nu}_{\mr{static}} & = \frac{16 \pi }{c^2} [m u^{\mu} u^{\nu}  - S^{j (\mu}u^{\nu)}\partial_{j} \\& -\frac{2}{3}J_{\mr{Q}}^{k (\mu \nu) l} \partial_{k}\partial_{l}  +  \frac{1}{3}J_{\mr{O}}^{kj(\mu \nu)i} \partial_{i}\partial_{j}\partial_{k}]\delta^{(3)}(x), \nonumber
\end{alignat}
where the indices $i,~j$ run over $1,2,3$ whereas $\mu,~\nu$ run over $0,1,2,3$.
In 3-D Fourier space, the solution to this is simply 
\begin{alignat}{3}
\label{static}
\tilde{h}^{\mu\nu}_{\mr{static}} &= \frac{-16 \pi}{c^2\vec{q}^2}[m u^{\mu} u^{\nu} + i q_j S^{j(\mu}u^{\nu)}\nonumber \\& +q_k q_l \frac{2}{3} J_{\mr{Q}}^{k (\mu \nu) l} + i q_i q_j q_k \frac{1}{3}J_{\mr{O}}^{kj(\mu \nu) i}],
\end{alignat}
where, $\tilde{h}^{\mu\nu}_{\mr{static}} \equiv \int d^3\vec{x} \exp[-i\vec{q}\cdot \vec{x}] h^{\mu\nu}(\vec{x})$ and  $i = \sqrt{-1}$.
We can Fourier invert Eq.~(\ref{static}) to find the expression in position space which gives 
\begin{alignat}{3}
\label{staticp}
h^{\mu\nu}_{\mr{static}} &= -\frac{4}{c^2} [m u^{\mu} u^{\nu}\big(\frac{1}{r}\big)  - S^{j (\mu}u^{\nu)}\partial_{j}\big(\frac{1}{r}\big) \\& -\frac{2}{3}J_{\mr{Q}}^{k (\mu \nu) l} \partial_{k}\partial_{l}\big(\frac{1}{r}\big)  +  \frac{1}{3}J_{\mr{O}}^{kj(\mu \nu)i} \partial_{i}\partial_{j}\partial_{k}\big(\frac{1}{r}\big)].  \nonumber
\end{alignat}

Substituting the expressions for the multipole moments, and switching to the use of the mass-normalized Pauli-Lubanski spin vector $a^{\mu} = -(1/2m)\epsilon^{\mu}_{~\nu\rho\sigma}u^{\nu}S^{\rho\sigma}$  we get 
\begin{alignat}{3}
h^{\mu\nu}_{\mr{st.}}&=-u^{\mu}u^{\nu}[1-\frac{C_2}{2}(a\cdot \partial)^2]\frac{4GM}{r} \nonumber \\& -u^{(\mu}\epsilon^{\nu)}{}_{\rho\alpha\beta}u^{\rho}a^{\alpha}\partial^{\beta}[1-\frac{C_3}{3!}(a\cdot \partial)^2]\frac{4GM}{r} \nonumber \\& + \text{terms containing }(\partial^2(1/r)=\delta^{(3)}(x)),
\end{alignat}
where we can neglect the delta function corrections since they only affect the metric at the location of the particle, where the field is anyway ill-defined. These terms essentially come from the non-zero traces of the quadrupole and octupole tensors in Eqs.~(\ref{quad}, \ref{octu}). As mentioned below  Eqs.~(\ref{quad}, \ref{octu}), they do not contribute to the final result.

Comparing the above metric perturbation (after dropping the delta function terms) to the linearized Kerr metric (see Ref.~\cite{Vines:2017hyw}), we can fix the coefficients to $C_2=C_3=1$ for the case of a black hole. However, we will continue to work with generic $C_2$ and $C_3$ for the rest of this work as it may be useful to have Compton amplitudes for generic compact bodies (satisfying parity-symmetry, stationarity and axisymmetry).

\subsection{Incident wave}
\label{perturber}
To compute the gravitational Compton amplitude, corresponding to the process of gravitational waves scattering off a black hole, we subject the black hole to an incoming gravitational wave. Although a realistic setup would involve a localized wave packet that eventually reaches the black hole and scatters off (and partially absorbed), we treat this situation in the manner of time-independent perturbation theory, and instead consider a monochromatic plane wave that had always been present. The monochromatic plane wave adds an additional metric perturbation ($h^{\mu\nu} = \sqrt{-g}g^{\mu\nu} -\eta^{\mu\nu}$) which we characterize as\footnote{Note that the metric perturbation should be real, and not a complex function but it is fine to work with $\exp[-ik\cdot x]$ as long as we truncate at linear order in $\epsilon$.}
\begin{alignat}{3}
\label{waveform}
& h_{\mr{w}}^{\mu \nu} = \epsilon ~ \varepsilon^{\mu} \varepsilon^{\nu} \exp[i k \cdot x] , ~ \nonumber \varepsilon^{\mu}\varepsilon_{\mu} = 0, ~\\& \varepsilon^{\mu} u_{\mu}^{(0)}=0, \quad \varepsilon^{\mu} k_{\mu} = 0,\quad k^{\mu} u_{\mu} = - \omega
\end{alignat}
where $\epsilon$ is the field strength and $\varepsilon^{\mu}$ is a  complex null vector. The conditions ensure that $h^{\mu \nu}_{\mr{wave}}$ is spatial, transverse and traceless, in the rest frame of the unperturbed particle and propagates at the speed of light. In this way, only the relevant radiative degrees of freedom are included. There are only two linearly independent vectors we can choose for $\varepsilon^{\mu}$, which fixes the helicity/polarization of the incoming gravitational wave. In the initial rest frame of the particle,  choosing $\varepsilon_{+2}^{\mu} \equiv (1/\sqrt{2})(0,1,\pm i,0)$ fixes the wave in $\pm 2$ helicity respectively. In the rest frame of the particle, we also choose the wave to be propagating along the +ve 'z' direction, $k^{\mu} \equiv \omega(1,0,0,1)$.

Subject to this incident plane linearized gravitational wave, the black hole experiences the following metric tensor. 
\begin{alignat}{3}
\label{metcor}
& g^{\mu \nu} = \eta^{\mu \nu} +  h_{\mr{w}}^{\mu \nu}, \quad g_{\mu \nu} = \eta_{\mu \nu} - h_{\mr{w},\mu \nu},\nonumber \\&~\sqrt{-g}=1+ \mathcal{O}(\epsilon^2).
\end{alignat}
The leading order Christoffel connection and the curvature due to the wave are given by
\begin{alignat}{3}
\label{wig}
\Gamma^{\rho}_{\mu\nu} &= \frac{-i \epsilon}{2}(\varepsilon^{\rho}\varepsilon_{\nu}k_{\mu}+\varepsilon^{\rho}\varepsilon_{\mu}k_{\nu} - k^{\rho}\varepsilon_{\nu}\varepsilon_{\mu})\exp[i k\cdot x]\nonumber \\&= \epsilon\exp[ik\cdot x]~ \Delta \Gamma^{\rho}_{\mu\nu},\\ ~ R^{\mu}_{~\nu \alpha \beta} &=\partial_{\alpha}\Gamma^{\mu}_{\nu \beta} - \partial_{\beta}\Gamma^{\mu}_{\nu \alpha} = \epsilon \exp[ik\cdot x]~ \Delta R^{\mu}_{~\nu\alpha\beta},\nonumber \\&
\label{wic}
\end{alignat}
where we have defined $\Delta \Gamma^{\rho}_{\mu\nu}$ and $\Delta R^{\mu}_{~\nu \alpha \beta}$ as the leading order in $\epsilon$ corrections to Christoffel connection after removing the factor $\epsilon\exp[ik\cdot x]$ for later convenience. The particle's dynamics (momentum, spin, etc.) are affected by the incident wave, subsequently perturbing stress energy tensor given in Eq.~(\ref{stress}) in an oscillatory fashion. We study the dynamics of the particle subject to this weak perturbation in the next section.

\section{Gravitational wave perturbation to the black hole's motion, spin and stress energy tensor}
\label{corrall}
\subsection{Solving for momentum, spin and 4-velocity}
\label{dynamics}
As mentioned earlier in Sec.~\ref{motion}, the equations of motion in Eqs.~(\ref{mots}, \ref{motp}) are incomplete. They are completed by the spin supplementary condition (SSC) $S^{\mu\nu}p_{\nu}=0$ and the normalization condition $u^{\mu}u_{\mu}=-1$. Further, the 4-velocity $u^{\mu}$ appears in the spin equation of motion (see Eq.~(\ref{mots})), and thus they cannot be solved independently. We need to identify the relation between $p^{\mu}$ and $u^{\mu}$ to solve for the spin. For that, we act upon the SSC with a ($D/D\tau$) and get\begin{alignat}{3}
 \frac{D S^{\mu\nu}}{D\tau} p_{\nu}& + S^{\mu\nu} \frac{D p_{\nu}}{D\tau}  = 0 = 2 p^{[\mu}u^{\nu]}p_{\nu}+N^{\mu\nu}p_{\nu}\nonumber \\& + S^{\mu\nu}F_{\nu}-\frac{1}{2}R_{\nu\alpha\rho\sigma}u^{\alpha}S^{\rho\sigma}S^{\mu\nu} ,
\\ & = p^{\mu} (u\cdot p) + m^2 u^{\mu} +N^{\mu\nu} p_{\nu} + S^{\mu\nu}F_{\nu}\nonumber \\&-\frac{1}{2}R_{\nu\alpha\rho\sigma} u^{\alpha}S^{\rho\sigma}S^{\mu\nu}, \\
-u^{\mu}  &= p^{\mu}\frac{u\cdot p}{m^2} + \frac{1}{m^2}(N^{\mu\nu}p_{\nu}+S^{\mu\nu}F_{\nu})\nonumber\\& -\frac{1}{2m^2}R_{\nu\alpha\rho\sigma}u^{\alpha}S^{\rho\sigma}S^{\mu\nu}, \label{4vel}
\end{alignat}
where $F^{\mu}$ and $N^{\mu\nu}$ were defined earlier in Eqs.~(\ref{mots}, \ref{motp}). Now, contracting with $u_{\mu}$ on both sides and using $p^{\mu}= m u^{\mu} + O(R)$ (i.e., momentum and 4-velocity are parallel in absence of external curvature) and $u^{\mu}u_{\mu}=-1$, we get
\begin{alignat}{3}
1 = \frac{(u.p)^2}{m^2} + O(R^2).
\end{alignat}
Note that the $m$ that appears here is not a fixed parameter, but a time-dependent mass function defined as $p^2=-m^2$. We can compute the mass function as follows,
\begin{alignat}{3}
&  -2 p\cdot \frac{Dp}{d\tau} = -\frac{Dp^2}{d\tau} = \frac{dm^2}{d\tau} \\& \approx m \frac{D}{D\tau}\bigg(\frac{1}{3}J_{\mr{Q}}^{\lambda\nu\rho\sigma} R_{\lambda\nu\rho\sigma} + \frac{1}{6}J_{\mr{O}}^{\tau\lambda\nu\rho\sigma}\nabla_{\tau}R_{\lambda\nu\rho\sigma}\bigg),\nonumber  \\
&\implies m = M +\frac{1}{6}J_{\mr{Q}}^{\lambda\nu\rho\sigma} R_{\lambda\nu\rho\sigma} + \frac{1}{12}J_{\mr{O}}^{\tau\lambda\nu\rho\sigma}\nabla_{\tau}R_{\lambda\nu\rho\sigma},\nonumber
\end{alignat}
where $M$ is the constant of integration to be interpreted as the constant mass in absence of spin-induced multipole moments and curvature. We also note that $m$ can be treated as a constant up to $S^2$. In the absence of curvature, or upto second order in spin, we can also write $p^{\mu} = M u^{\mu}$.

Now, substituting the expression for force $F^{\mu}$, and torque $N^{\mu\nu}$, from Eqs.~(\ref{mots}, \ref{motp}), in Eq.~(\ref{4vel}) and simplifying, we get
\begin{alignat}{3}
\label{4vels}
u^{\mu} &= \frac{p^{\mu}}{m} + \frac{1}{m^2}(N^{\mu\nu}p_{\nu}+S^{\mu\nu}F_{\nu})\\& -\frac{1}{2m^2}R_{\nu\alpha\rho\sigma}u^{\alpha}S^{\rho\sigma}S^{\mu\nu}.
\end{alignat}
The commutator $p^{[\mu}u^{\mu]}$ that enters the expression for $DS^{\mu\nu}/D\tau$ (see Eq.~(\ref{mots})) can now be computed. 
Thus, we have now assembled all the expressions needed to solve for the perturbations to the 4-velocity, momentum and spin of the particle due to the incident gravitational wave perturbation.
\subsection{Correction to momentum, spin and spin-induced multipole moments}
\label{corrections}
The gravitational perturbation is an incident plane wave, that goes as $~\exp[ik\cdot x]\approx \exp[-i\omega \tau]$ at the particle's location. Thus, we anticipate that at linear order in wave strength $\epsilon$, the dynamical variables are going to be affected in an oscillatory fashion, and expand them as 
\begin{alignat}{3}
\label{expp}
p^{\mu} &= p^{(0)\mu} + \epsilon \exp[-i\omega \tau]~\Delta p^{\mu}, \\
\label{expu}
u^{\mu} &= u^{(0)\mu} + \epsilon \exp[-i\omega \tau]~\Delta u^{\mu}, \\
\label{exps}
S^{\mu\nu} &= S^{(0)\mu\nu} + \epsilon \exp[-i\omega \tau]~\Delta S^{\mu\nu},
\end{alignat}
where $\epsilon$ is the strength of the wave, and $u^{(0)\mu}\equiv(1,0,0,0)$, $	p^{(0)\mu} = m u^{(0)\mu}$, and $S^{(0)\mu\nu}$ are the undisturbed original constant values in the absence of the external perturbation. The correction to the momentum can be easily solved by substituting the expansions in Eqs.~(\ref{expp}, \ref{wig}, \ref{wic}) into Eq.~(\ref{motp}) which gives
\begin{alignat}{3}
\label{corp}
& -i\omega\Delta p_{\mu} = + M \Delta \Gamma^{\nu}_{\mu\gamma} u^{(0)\gamma} u^{(0)}_{\nu}-\frac{1}{2} u^{\nu}S^{\rho\sigma} \Delta R_{\mu\nu\rho\sigma}\nonumber \\& - \frac{i}{6} J_{\mr{Q}}^{(0)\lambda\nu\rho\sigma}k_{\mu}\Delta R_{\lambda\nu\rho\sigma} + \frac{1}{12} J_{\mr{O}}^{(0)\tau\lambda\nu\rho\sigma}k_{\mu}k_{\tau}\Delta R_{\lambda\nu\rho\sigma}, \nonumber \\&
\end{alignat}
where $\Delta R_{\lambda\nu\rho\sigma}$ is proportional to the curvature perturbation due to the incident wave defined earlier in Eq.~(\ref{wic}).
The connection term arises from expanding the covariant derivative. The correction to the 4-velocity can be easily obtained by taking the variation of either Eq.~(\ref{4vel}) or Eq.~(\ref{4vels}) (here we use the former expression), which gives
\begin{alignat}{3}
\label{coru}
\Delta u^{\mu} &= \Delta(p^{\mu}/m)-\frac{1}{i\omega M^2}(M N^{\mu\nu}p^{(0)}_{\nu}+S^{(0)\mu\nu}F_{\nu}) \nonumber \\&-\frac{1}{2 M^2}\Delta R_{\nu\alpha\rho\sigma}u^{(0)\alpha}S^{(0)\rho\sigma}S^{(0)\mu\nu}.
\end{alignat}

Finally, we can solve for the spin by taking the variation  on both sides of Eq.~(\ref{mots}) to get
\begin{alignat}{3}
\label{cors}
\nonumber -i\omega \Delta S^{\mu\nu} &=-2u^{(0)\lambda}S^{(0)\delta[\nu}\Delta\Gamma^{\mu]}_{\lambda\delta} -2 i\omega  \Delta(p^{[\mu} u^{\nu]})\\& + \frac{4}{3}\Delta R^{[\mu}{}_{\lambda\rho\sigma}J_{Q}^{(0)\nu]\lambda\rho\sigma} + \frac{2}{3}i k^{\lambda}\Delta R^{[\mu}{}_{~\tau\rho\sigma}J_{\mr{O}\lambda}^{(0)\nu]\tau\rho\sigma}\nonumber \\&+\frac{i k^{[\mu}}{6}\Delta R_{\lambda\tau\rho\sigma}J_{\mr{O}}^{(0)\nu]\lambda\tau\rho\sigma},
\end{alignat}
where $\nabla \Gamma^{\mu}_{\nu\delta}$ is proportional to the Christoffel connection defined in Eq.~(\ref{wig}).
The correction to the worldline can be obtained from the correction to the 4-velocity as
\begin{alignat}{3}
\label{expz}
z^{\mu}(\tau) &= \int d\tau u^{\mu}(\tau) = \int d\tau [u^{(0)}  + \epsilon \exp[-i \omega \tau] \Delta u^{\mu}] \\& = u^{(0)}\tau - \frac{1}{i\omega}\epsilon \exp[-i \omega\tau] \Delta u^{\mu}.
\end{alignat}

The correction to the spin-induced multipole moments can be obtained by substituting into their expressions into Eqs.~(\ref{quad}, \ref{octu}), the expansions in Eqs.~(\ref{expp}, \ref{expu}, \ref{exps}) (for $p$, $u$, and $S$) and  Eq.~(\ref{metcor}) (as the metric enters the expressions for multipole moments through contractions). The explicit expressions are not very illuminating but it is useful to separate the two ways in which the multipole moments are perturbed. We first rewrite the expressions for the multipole moments in Eqs.~(\ref{quad}, \ref{octu}) as 
\begin{alignat}{3}
J_{\mr{Q}}^{\mu\nu\rho\sigma} &= \frac{3C_2}{m} u^{[\mu} S^{\nu]\lambda}g_{\lambda\alpha}S^{\alpha[\rho}u^{\sigma]},\\  J_{\mr{O}}^{\lambda \mu \nu \rho\sigma} &= \frac{C_3}{4 m^2}[\Theta^{\lambda[\mu}u^{\nu]}S^{\rho\sigma} + \Theta^{\lambda[\rho}u^{\sigma]}S^{\mu\nu}-\Theta^{\lambda[\mu}S^{\nu]\rho}u^{\sigma}\nonumber \\& -\Theta^{\lambda[\rho}S^{\sigma][\mu}u^{\nu]}-S^{\lambda[\mu}\Theta^{\nu][\rho}u^{\sigma]}-S^{\lambda[\rho}\Theta^{\sigma][\mu}u^{\nu]}],
\nonumber\\ \quad & \Theta^{\mu\nu}= S^{\mu \lambda}S^{\nu\alpha}g_{\lambda\alpha},
\end{alignat} 
and define 
\begin{alignat}{3}
\label{corq}
\Delta J_{Q,\mr{contact}}^{\mu\nu\rho\sigma} &= -\frac{\partial J_{Q}^{\mu\nu\rho\sigma}}{\partial g_{\alpha\beta}}  \varepsilon_{\alpha}\varepsilon_{\beta},   \\ \Delta J_{Q,\mr{matter}}^{\mu\nu\rho\sigma} &= \frac{\partial J_{Q}^{\mu\nu\rho\sigma}}{\partial S^{\alpha\beta}} \Delta S^{\alpha\beta}  + \frac{\partial J_{Q}^{\mu\nu\rho\sigma}}{\partial u^{\alpha}} \Delta u^{\alpha}, \\
\Delta J_{O,\mr{contact}}^{\lambda\mu\nu\rho\sigma} &= -\frac{\partial J_{O}^{\lambda\mu\nu\rho\sigma}}{\partial g_{\alpha\beta}}  \varepsilon_{\alpha}\varepsilon_{\beta},   \\ \Delta J_{O,\mr{matter}}^{\lambda\mu\nu\rho\sigma} &= \frac{\partial J_{O}^{\lambda\mu\nu\rho\sigma}}{\partial S^{\alpha\beta}} \Delta S^{\alpha\beta}  + \frac{\partial J_{O}^{\lambda\mu\nu\rho\sigma}}{\partial u^{\alpha}} \Delta u^{\alpha}. \nonumber \\&
\label{coro}
\end{alignat}
The quantities with subscript ``contact'' are corrections to the multipole moments through the explicit dependence on metric tensor (after removing the factor $\epsilon \exp[-i\omega \tau]$). The quantities with subscript ``matter'' are corrections acquired through the corrections to 4-velocity and spin tensor (variables describing the dynamics of matter). This separation is gauge dependent and purely for our convenience. With these conventions, the expressions for the perturbed spin-induced multipole moments are given by
\begin{alignat}{3}
\label{expqq}
J_Q^{\mu\nu\rho\sigma} &= J_Q^{(0)\mu\nu\rho\sigma} \\& \nonumber +\epsilon \exp[-i\omega \tau](\Delta J_{Q,\mr{contact}}^{\mu\nu\rho\sigma}+\Delta J_{Q,\mr{matter}}^{\mu\nu\rho\sigma})  \\& \nonumber \\  \label{expoo}
J_O^{\mu\nu\rho\sigma} &= J_O^{(0)\mu\nu\rho\sigma} \\&  +\epsilon \exp[-i\omega \tau](\Delta J_{O,\mr{contact}}^{\mu\nu\rho\sigma}+\Delta J_{O,\mr{matter}}^{\mu\nu\rho\sigma}) \nonumber
\end{alignat}
\subsection{Correction to the stress energy tensor}

Finally, equipped with the perturbations to momentum, spin, 4-velocity and the worldline (Eqs.~(\ref{corp}, \ref{coru}, \ref{cors}, \ref{expz})), and the spin-induced multipole moments (Eqs.~(\ref{corq}, \ref{coro})), we can compute the perturbation to the stress energy tensor of the particle. This is accomplished by simply substituting the expansions in Eqs.~(\ref{expu}, \ref{expp}, \ref{exps}, \ref{expqq}, \ref{expoo}) and expressions for curvature tensor and connection coefficients given in Eqs.~(\ref{wic}, \ref{wig}) into the expression for the stress energy tensor given in Eq.~(\ref{stress}) and then truncating at leading order in $\epsilon$. We also discard any $O(S^4)$ contributions.

The explicit expressions for the corrections to the stress energy tensor are quite complicated and thus will not be written here, but we find it convenient to separate them into pieces in a manner similar to how we separated the corrections to the multipole moments (see Eqs.~(\ref{corq}, \ref{coro})). The stress energy tensor too gains corrections from two sources: (i) from the correction to the dynamical variables describing the particle (matter content, $p$,  $u$, $S$, $z$) and (ii) from the explicit dependence on the curvature tensor (and derivatives), Christoffel connection terms (from the covariant derivatives), and on the metric tensor (that enters the multipole moments) in Eq.~(\ref{stress}).
The latter contributes even if the particle's motion and dynamics are unaffected. Thus, after expanding out the covariant derivatives in the stress energy tensor in Eq.~(\ref{stress}), we define
\begin{widetext}
\begin{alignat}{3}
\label{constress}
\Delta T^{\mu\nu}_{\mr{contact}}(x) &= \frac{\partial T^{\mu\nu}}{\partial R^{\alpha \beta \gamma \delta}} \Delta R^{\alpha\beta\gamma\delta}\exp[ik \cdot x] +\frac{\partial T^{\mu\nu}}{\partial (\partial_{\lambda}R^{\alpha \beta \gamma \delta)}} \Delta R^{\alpha\beta\gamma\delta}i k_{\lambda}\exp[ik \cdot x] \\ \nonumber & - \frac{\partial T^{\mu\nu}}{\partial(\partial_{\lambda}\partial_{\rho} R^{\alpha \beta \gamma \delta})}  \Delta R^{\alpha\beta\gamma\delta}k_{\lambda}k_{\rho}\exp[ik \cdot x] + \frac{\partial T^{\mu\nu}}{\partial\Gamma^{\alpha}_{\beta\gamma}}\Delta \Gamma^{\alpha}_{\beta\gamma}  \exp[ik\cdot x] \\ &+ \frac{\partial T^{\mu\nu}}{\partial J^{\alpha\beta\gamma\delta}_{Q,\mr{contact}}}\Delta J^{\alpha\beta\gamma\delta}_{Q,\mr{contact}}  \exp[-i\omega \tau] +  \frac{\partial T^{\mu\nu}}{\partial J^{\tau\alpha\beta\gamma\delta}_{O,\mr{contact}}}\Delta J^{\tau\alpha\beta\gamma\delta}_{O,\mr{contact}} \exp[-i\omega \tau] \nonumber
\\ \nonumber \\ \label{matstress}\Delta T^{\mu\nu}_{\mr{matter}}(x) &= \frac{\partial T^{\mu\nu}}{\partial p^{\alpha}}\Delta p^{\alpha} \exp[-i\omega \tau] + \frac{\partial T^{\mu\nu}}{\partial  S_{\alpha \beta}}\Delta S_{\alpha\beta}\exp[-i\omega \tau] + \frac{\partial T^{\mu\nu}}{\partial u^{\alpha}}\Delta u^{\alpha} \exp[-i\omega \tau]\\& + \frac{\partial T^{\mu\nu}}{\partial z^{\alpha}}\Delta z^{\alpha}  \exp[-i\omega \tau] + \frac{\partial T^{\mu\nu}}{\partial J^{\alpha\beta\gamma\delta}_{Q,\mr{matter}}}\Delta J^{\alpha\beta\gamma\delta}_{Q,\mr{matter}}  \exp[-i\omega \tau] +  \frac{\partial T^{\mu\nu}}{\partial J^{\tau\alpha\beta\gamma\delta}_{O,\mr{matter}}}\Delta J^{\tau\alpha\beta\gamma\delta}_{O,\mr{matter}} \ \exp[-i\omega \tau] \nonumber
\end{alignat}
\end{widetext}
 Once again, the idea is to separate the perturbation due to matter/dynamical variables ($p$, $S$, $u$, $z$) and that due to explicit corrections to metric, connection and Christoffel coefficients. Note that there is no difference between $\epsilon \exp[ik\cdot x]$ and $\epsilon \exp[-i\omega \tau]$ as all terms in the stress energy tensor contain $\delta^{(4)}(x-z(\tau))$ or its derivatives. Thus, as $z^{\mu}(\tau) = u^{\mu}\tau + O(\epsilon)$, $\epsilon \exp[ik\cdot x] = \epsilon \exp[-i\omega \tau] + O(\epsilon^2)$.

 With these conventions, the perturbed stress energy tensor is thus given by
 \begin{alignat}{3}
 T^{\mu\nu}(x) &= T^{(0)\mu\nu} + \epsilon\Delta T^{\mu\nu}\\& =  T^{(0)\mu\nu}(x) +\epsilon \Delta T^{\mu\nu}_{\mr{matter}}(x)+\epsilon\Delta T^{\mu\nu}_{\mr{contact}}(x), \nonumber
 \end{alignat}
 where $T^{(0)\mu\nu}(x)$ is the unperturbed stress energy tensor of the free particle given in Eq.~(\ref{flatstress}). The correction terms to the stress energy tensor consist of terms of the form 
 \begin{alignat}{3}
 \label{structure}
 \Delta T^{\mu\nu}_{m/c} =& \sum \int d\tau A_{m/c}^{\mu\nu K L}[e^{\alpha},S^{(0)\alpha\beta},u^{(0)\alpha},m]\nonumber \\\times~ &  \partial_{K}[\exp[ik\cdot x]\partial_{L}\delta^{(4)}(x-u^{(0)\mu}\tau)],
 \end{alignat}
 where we are using the multi-index notation $L = \mu_1\mu_2...\mu_l$, to denote a chain of indices. The $\sum$ shows that the net result is a linear combination of such terms (with different $l$, $k$ and $A^{\mu\nu}$). $m/c$ refers to ``matter/contact'', both types of corrections have the same form. 
 We now proceed to compute the scattering amplitude by solving the Einstein equation for the scattered wave.

\section{Computation of the scattering amplitude}
\label{scwan}
\subsection{Differential equation for the scattered wave}
To derive a differential equation governing the scattered wave, we first rewrite the Einstein equation in Landau-Lifshitz form (see Ref.~\cite{Blanchet:2013haa}), i.e.,
\begin{subequations}
\label{eeq}
\begin{alignat}{3}
& \eta^{\rho\sigma}\partial_{\rho}\partial_{\sigma}h^{\alpha \beta} = 16 \pi G \tau^{\alpha \beta}, \\
& \tau^{\alpha \beta} = |g| T^{\alpha \beta} + \frac{1}{16 \pi G} \Lambda^{\alpha \beta}, \\
\Lambda^{\alpha \beta} &= -h^{\mu \nu}\partial^2_{\mu\nu}h^{\alpha \beta} + \partial_{\mu}h^{\alpha \nu}\partial_{\nu}h^{\beta \mu} \nonumber \\& + \frac{1}{2}g^{\alpha \beta}g_{\mu\nu}\partial_{\lambda}h^{\mu \tau}\partial_{\tau}h^{\nu \lambda}- g^{\alpha \mu}g_{\nu \tau}\partial_{\lambda}h^{\beta \tau}\partial_{\mu}h^{\nu \lambda}\nnm \\&-g^{\beta \mu}g_{\nu \tau}\partial_{\lambda}h^{\alpha \tau}\partial_{\mu}h^{\nu \lambda}+g_{\mu \nu}g^{\lambda \tau}\partial_{\lambda}h^{\alpha \mu}\partial_{\tau}h^{\beta \nu} \nnm \\&+\frac{1}{8}(2g^{\alpha\mu}g^{\beta \nu}- g ^{\alpha \beta}g^{\mu \nu})(2 g_{\lambda \tau} g_{\epsilon \pi} - g_{\tau \epsilon} g_{\lambda \pi})\nnm\\&\times\partial_{\mu} h^{\lambda \pi} \partial_{\nu} h^{\tau \epsilon}, \\
h^{\mu\nu} &= \sqrt{-g}g^{\mu\nu} -\eta^{\mu\nu}.
\end{alignat}
\end{subequations}
We then substitute 
\begin{alignat}{3}
\sqrt{-g}g^{\mu \nu} - \eta^{\mu \nu} = h^{\mu\nu}&=  G h^{\mu \nu}_{\mr{static}} + \epsilon ~ \varepsilon^{\mu} \varepsilon^{\nu} ~\exp[i k \cdot x]  \nnm \\& + G \epsilon  h_{\mr{scatter}}^{\mu \nu},
\end{alignat}
anticipating the scaling of the various metric perturbations.
Here, $h^{\mu\nu}_{\mr{static}}$ is the static metric correction given in Eq.~(\ref{staticp}), sourced by the compact body and $\propto G$. The second term is the incident wave given earlier in Eq.~(\ref{waveform}) and $\propto \epsilon$. Finally, the last term is the scattered wave which starts at leading order in $G\epsilon$. 
To isolate the leading order (in $G$) scattered wave, we take the coefficient of $G\epsilon$ in both sides of Eq.~(\ref{eeq}) after substitution. This gives us
\begin{alignat}{3}
\label{sceqnform}
\eta^{\rho\sigma}\partial_{\rho}\partial_{\sigma} h_{\mr{scatter}}^{\mu\nu}(x)  &=16 \pi \Delta T^{\mu\nu}_{\mr{matter}}(x) + 16 \pi \Delta T^{\mu\nu}_{\mr{contact}}(x)\nonumber \\& + \Lambda^{(1,1),\mu\nu}(x),
\end{alignat}
where $\Delta T^{\mu\nu}_{\mr{matter},\mr{contact}}$ are the leading order (in $\epsilon$) perturbations to the stress energy tensor of the particle due to the incident plane wave defined in Eqs.~(\ref{constress}, \ref{matstress}). $\Lambda^{(1,1),\mu\nu}$ is the term proportional to $G\epsilon$ when we substitute $\sqrt{-g}g^{\mu \nu} - \eta^{\mu \nu} = G h^{\mu \nu}_{\mr{static}} + \epsilon~ \varepsilon^{\mu} \varepsilon^{\nu}\exp(i k \cdot x)$ in $\Lambda^{\mu\nu}$ \footnote{Thus, we are not interested in the non linear interactions proportional to $G^2$ or $\epsilon^2$ in this work. They do not contribute to the leading order scattered wave which goes as $G\epsilon$.}. This is the contribution from non linear interactions in gravity, between the incident wave and the static curvature. Thus the scattered wave obeys a wave equation with a source term derived from three contributions,a ``matter'' part from $\Delta T^{\mu\nu}_{\mr{matter}}$  which comes from the perturbing the dynamics ($p^{\mu}$, $S^{\mu\nu}$) and kinematics ($u^{\mu}$, $z^{\mu}$) of the particle, a ``contact'' part from $\Delta T^{\mu\nu}_{\mr{contact}}$ which is due to the explicit dependence of the stress energy tensor on the metric and its derivatives. This also includes the corrections to stress energy tensor arising from the explicit metric dependence in the expression for multipole moments.
The last part comes from $\Lambda^{(1,1)\mu\nu}$ is due to the non linear interactions in gravity, wherein the incident wave scatters off the static linearized gravitational field sourced by the particle. We refer to the last contribution as the ``graviton'' part. This separation is primarily for convenience and is gauge dependent, i.e., they mix with each other under coordinate transformations.

\subsection{Solving for the scattered wave}
\label{child}

It is more convenient to solve the differential equation.~(\ref{sceqnform}) for the scattered wave in Fourier space. Furthermore, we are only interested in the radiative on-shell modes whose mode-vectors $l^{\mu}$ satisfy the null condition ($l^2=0$). In Fourier space, Eq.~(\ref{sceqnform}) becomes
\begin{alignat}{3}
\label{ftsceqn}
- l^{2} \tilde{h}^{\mu\nu}_{\mr{scatter}}(l) &=  \mathcal{F}[16\pi \Delta{T}^{\mu\nu}_{\mr{matter}}(x) + 16 \pi \Delta T^{\mu\nu}_{\mr{contact}}(x) \nonumber \\& + \Lambda^{(1,1),\mu\nu}(x)](l),
\end{alignat}
where $\mathcal{F}$ is the Fourier transform operator defined as $\mathcal{F}(f(x)) = \int d^4{x} \exp[-il\cdot x]f(x)$. 

The form of Fourier transform of $\Delta T^{\mu\nu}_{\mr{matter/contact}}(x)$ can be evaluated using the fact that both of these corrections consist of terms of the form given in Eq.~(\ref{structure}) as shown below.
\begin{subequations}
\label{ftmorc}
\begin{alignat}{3}
&\mathcal{F}(\Delta T^{\mu\nu}_{\mr{m/c}})(l)= \int d\tau d^4x \exp[-il\cdot x]\Delta T^{\mu\nu}_{\mr{m/c}}(x) \nnm
 \\\nnm
 =& \sum A_{m/c}^{\mu\nu K L}\int \exp[-il \cdot x]\prod_{i=1}^{k} \partial_{\mu_i} \exp[ik\cdot x] \\& \times  \prod_{j=1}^{l} \partial_{\mu_j} \delta^{(4)}(x-u^{(0)\mu}\tau), \\
=& \sum A_{m/c}^{\mu\nu K L}(i)^{l}\int d\tau d^4x \prod_{i=1}^{k} l_{\mu_i} \exp[-i(l-k)\cdot x]\nnm \\& \times\prod_{j=1}^{l} \partial_{\mu_j} \delta^{(4)}(x-u^{(0)\mu}\tau),  \\
=& \sum(i)^n A_{m/c}^{\mu\nu K L} \int d\tau d^4 x \prod_{i=1}^{k} l_{\mu_i} \prod_{j=1}^{l}(l-k)_{\mu_j} \\& \times \exp[i(l-k)\cdot x]\delta^{(4)}(x-u^{(0)\mu}\tau), \nonumber\\
=& \sum(i)^n 2\pi A_{m/c}^{\mu\nu K L}  l_{K} (l-k)_{L}\delta(l\cdot u^{(0)}-k\cdot u^{(0)}) \nonumber \\=& \sum(i)^n 2\pi A_{m/c}^{\mu\nu K L}  l_{K} (l-k)_{L}\delta(l\cdot u^{(0)}+\omega),
\end{alignat}
\end{subequations}
where we see that the frequency preserving delta function $\delta(l\cdot u^{(0)}+\omega)$ ensures that the outgoing wave has the same frequency as the incoming wave. We have included the $\sum$ operator to indicate that the net result is a sum of multiple such terms.

The form of the Fourier transform of the ``graviton'' part from $\Lambda^{(1,1)}$ can be evaluated using the fact that it consists of sum of products of the static metric perturbation sourced by the particle given in Eq.~(\ref{static}) and the incident wave described in Eq.~(\ref{waveform}). Thus, $\Lambda^{(1,1)}$ is a sum of terms of the form $A_{\lambda}^{\mu\nu K}\exp[ik\cdot x] \times \partial_{K}\big(\frac{1}{r}\big)$, in the rest frame of the particle, and we have once again used the multi-index notation $K = \mu_1\mu_2...\mu_k$. We have
\begin{subequations}
\begin{alignat}{3}
& \mathcal{F}(\Lambda^{(1,1)}(x)) \sim \sum \int d^4 x A_{\lambda}^{\mu\nu K}\exp[-il\cdot x][\exp[ik\cdot x] \nnm \\& \times \quad \prod_{i=1}^{k}\partial_{\mu_i}\bigg(\frac{1}{r}\bigg)]  \\& = \sum \int A_{\lambda}^{\mu\nu K} d^4 x \exp[-i(l-k)\cdot x]\prod_{i=1}^{k}\partial_{\mu_i}\bigg(\frac{1}{r}\bigg), \nonumber \\
&=\sum (i)^{n} \int d^4 x A_{\lambda}^{\mu\nu K} [(l-k)_{L}]\exp[i(l-k)\cdot x] \bigg(\frac{1}{r}\bigg) \nonumber \\ &= \sum A^{\mu\nu K}_{\lambda} (-1)^n (l-k)_{K} \frac{4\pi\delta(l^{0}-k^{0})}{(\vec{l}-\vec{k})^2},
\end{alignat}
\end{subequations}
in the rest frame of the particle, we can make this covariant by writing $l^{0} = l \cdot u^{(0)\mu}$ (similarly for $k$) and using $(\vec{l}-\vec{k})^2 = (l-k)^2 + O(G,\epsilon)$. Thus, we get 
\begin{alignat}{3}
\mathcal{F}(\Lambda^{(1,1)}(x)) = \sum  A_{\lambda}^{\mu\nu K}(l-k)_{K} \frac{4\pi\delta(l\cdot u^{(0)}+\omega)}{(l-k)^2}, \nnm \\ &
\label{ftgrav}
\end{alignat}
where once again we find the frequency preserving delta function.

Substituting the Fourier transforms of the matter, contact and graviton parts from Eqs.~(\ref{ftmorc}, \ref{ftgrav}) into Eq.~(\ref{ftsceqn}), we get the scattered wave in the Fourier domain with the form 
\begin{alignat}{3}
\tilde{h}^{\mu\nu}_{\mr{scatter}}(l^{\mu}) &= 
\frac{-1}{4\pi}\sum \Bigg\{[ A^{\mu\nu K L}_{m} +   A^{\mu\nu K L}_{c}](i)^n l_K (l-k)_L  \nnm \\& - 2 A_{\lambda}^{\mu\nu K} \frac{(l-k)_K}{(l-k)^2}\Bigg\}\frac{8 \pi^2 \delta(\omega + l\cdot u^{(0)})}{\omega l^2}\\ &= A^{\mu\nu}((l-k)^{\alpha},k^{\alpha}, \varepsilon^{\alpha}, S^{(0)\alpha\beta}, u^{(0)\alpha},m) \nnm \\&\times \frac{8\pi^2\delta(\omega + l\cdot u^{(0)})}{\omega l^2},
\label{finamp}
\end{alignat}
where the amplitude tensor $A^{\mu\nu}$ contains the scattering amplitudes for on-shell modes ($l^2\rightarrow 0$). Finally, we define the amplitudes as
\begin{alignat}{3}
\mathcal{M}_{\pm +} = G A^{\mu\nu}\zeta_{-2\nu}\zeta_{-2\nu},\mathcal{M}_{\pm -} = G A^{\mu\nu}\zeta_{+2\nu}\zeta_{+2\nu}, \nnm \\&
\label{amps}
\end{alignat}
where the first sign is fixed by the polarization of the incoming wave. The form of the wave in Fourier domain in Eq.~(\ref{finamp}) may seem a bit odd. It is much simpler to understand the form of the scattered wave in position space, where it is just a spherical wave with an angle dependent amplitude (in the rest frame of the particle). To see that, we will evaluate the Fourier transform of the scattered wave. We follow the procedure given in Ref.~\cite{Kosower:2018adc}. We first rewrite the differential equation in position space given in Eq.~(\ref{sceqnform}) as 
\begin{alignat}{3}
\Box h^{\mu\nu}_{\mr{scatter}}(x) = J^{\mu\nu}(x),
\end{alignat}
and write the solution as 
\begin{alignat}{3}
h^{\mu\nu}(x) &= \int d^4 x'~ G(x-x')J^{\mu\nu}(x'), \\ G(x)& =\frac{1}{2\pi}\theta(x^{(0)})\delta(x^2),
\end{alignat}
We now fourier transform only w.r.t.\ time in rest frame of BH (defined by $u^{\mu}$) to get,
\begin{alignat}{3}
h^{\mu\nu}(\omega,\vec{x}) &=\int dt \exp[i\omega t] h^{\mu\nu}(t,\vec{x}) \\& = \int dt' d^3 \vec{x}' \frac{\exp[i\omega (t'+|\vec{x}-\vec{x}'|)]}{4\pi |\vec{x}-\vec{x}'|}J^{\mu\nu}(t',\vec{x}'), \\&= \frac{1}{4\pi}\int d^3 x'  \frac{\exp(i\omega|\vec{x}-\vec{x}'|)}{|\vec{x}-\vec{x}'|}J^{\mu\nu}(\omega,\vec{x}'),
\end{alignat}
Far away from the black hole, we can approximate $|\vec{x}'|\ll |\vec{x}|$, and thus write $|\vec{x}-\vec{x}'|\approx |\vec{x}|-i\omega \hat{x}\cdot\vec{x}'$ and ignore the correction to $|\vec{x}-\vec{x}'|$ in the denominator to get
\begin{alignat}{3}
h^{\mu\nu}(\omega,\vec{x})&\approx \frac{\exp[i\omega |\vec{x}|]}{4\pi |\vec{x}|}\int d^3\vec{x}'\exp[-i\omega \hat{x}\cdot \vec{x}']J^{\mu\nu}(\omega,\vec{x}') \nnm \\& = \frac{\exp[i\omega |\vec{x}|]}{4\pi|\vec{x}|}\tilde{J}^{\mu\nu}(\bar{k}),
\end{alignat}
where $\tilde{J}^{\mu
\nu}(\bar{k})$ is the 4-D Fourier transform of $J^{\mu\nu}(x)$ evaluated at $\bar{k}=(\omega,\omega \hat{x})$.
Now, substituting the expression for $\tilde{J}^{\mu\nu}(\bar{k})$ from Eq.~(\ref{finamp}), we get 
\begin{alignat}{3}
h^{\mu\nu}(\omega,\vec{x}) = \frac{2\pi\exp[i\omega|\vec{x}|]}{\omega\vec{\vec{x}}}\delta(\omega+l\cdot u^{(0)})A^{\mu\nu}, 
\end{alignat}
and finally inverting the fourier transform along time, we get the result
\begin{alignat}{3}
h^{\mu\nu}(t,\vec{x}) = \frac{\exp[i\omega (r-t)]}{\omega r} A^{\mu\nu},~r=|\vec{x}|.
\end{alignat}
We find that the $A^{\mu\nu}$ defined in Eq.~(\ref{finamp}) is simply the amplitude of a spherical wave ($\exp[i\omega (r-t)]/(\omega r)$) in position space, centred at the particle. Further, using the definition of the scattering amplitudes as given in Eq.~(\ref{amps}), we can rewrite the relevant part of the scattered wave (the part that contributes to the amplitudes, $\mathcal{M}_{\pm\pm}$, as) 
\begin{alignat}{3}
h^{\mu\nu}(t,\vec{x})|_\tr{relevant} &= \frac{\exp[i\omega (r-t)]}{G \omega r} (\mathcal{M}_{\pm +} \zeta_{+2}^{\mu}\zeta_{+2}^{\nu}\nnm \\& + \mathcal{M}_{\pm -} \zeta_{-2}^{\mu}\zeta_{-2}^{\nu}),
\end{alignat}
where the first sign of the amplitudes is fixed by the polarization of the incoming wave.

\section{Compton amplitudes}
\label{result}
Finally, in this section, we project out the amplitude function $A^{\mu\nu}$ onto an appropriate set of polarization vectors and write down the scattering amplitudes to third order in spin, for generic $C_2$, $C_3$. To do that, we first fix the incident wave at a given helicity ($\pm 2$), by choosing the polarization vectors appropriately ($\varepsilon^{\mu}\equiv (1/\sqrt{2})(0,1,\pm i,0)$). Then, the scattering amplitude for measuring an outgoing wave with wave vector $l^{\mu}$, with $+ 2~(-2)$ helicity is given by $\mathcal{M}_{\pm+}=G A^{\mu\nu} \zeta_{-2,\mu}\zeta_{-2,\nu}$ [$\mathcal{M}_{\pm -}=G A^{\mu\nu} \zeta_{+2,\mu}\zeta_{+2,\nu}$] respectively,
where we are using the notation $\mathcal{M}_{\pm\pm}$ for the scattering amplitudes from a given incoming helicity to a given outgoing helicity. Here, $\zeta_{\pm2}^{\mu}$ are the complex null polarization vectors orthogonal to the outgoing wavevector $l^{\mu}$
(just as $\varepsilon^{\mu}$ are the complex null polarization vectors for the incoming wave with wavevector $k^\mu$), and they satisfy $\zeta_{a}\cdot \zeta_{b} = (1-\delta_{ab})$, $\zeta_a\cdot l = 0$. 

With this formula, we obtain the following expressions for the amplitudes for the BH case, i.e., $C_2=C_3=1$.
\begin{alignat}{3}
\label{amp2223m}
\mathcal{M}_{++} &= G M \omega \frac{\cos^4\big(\frac{\theta}{2}\big)}{\sin^2\big(\frac{\theta}{2}\big)} ~ \exp\bigg[{-}\frac{s^{\mu}}{M}(k_{\mu}+l_{\mu})\tan^2\big(\frac{\theta}{2}\big)\nnm\\& - \frac{i}{M \omega \cos^2(\frac{\theta
}{2})} S^{\mu\nu}k_{\mu}l_{\nu}\bigg]+O(S^4),\\\label{amp22m23m}
\mathcal{M}_{+-} &= G M \omega \sin^2\bigg(\frac{\theta}{2}\bigg)~\exp\bigg[\frac{s^{\mu}}{M}(k_{\mu}-l_{\mu})\bigg]+O(S^4),\nnm\\&\\ 
\label{ampm22m23m} \mathcal{M}_{--} &= \bar{\mathcal{M}}_{++}(S^{\mu\nu}\rightarrow -S^{\mu\nu}, ~s^{\mu}\rightarrow -s^{\mu}), \\
\label{ampm2223m}\mathcal{M}_{-+} &= \bar{\mathcal{M}}_{+-}(S^{\mu\nu}\rightarrow -S^{\mu\nu}, ~s^{\mu}\rightarrow -s^{\mu}),
\end{alignat}
where $k^{\mu}\equiv \omega(1,0,0,1)$ is the incoming wave vector and $l^{\mu} \equiv \omega(1,\hat{n})$ is the outgoing wave vector, with $\hat{n}$ showing the spatial direction and we are using the notation $\bar{f}$ to denote the complex conjugate of $f$. $\theta$ is the angle between incident wave-vector and outgoing wave-vector and related to the wave-vectors via $-k\cdot l = \omega^2 - \omega^2 \cos(\theta) = 2\omega^2 \sin^2(\frac{\theta}{2})$. $s^{\mu} = -(1/2)\epsilon^{\mu}{}_{\nu\rho\sigma}u^{\nu}S^{\rho\sigma}$ is the  Pauli-Lubanksi spin vector. The spin $S^{\mu\nu}=S^{(0)\mu\nu}$ and 4-velocity $u^{\mu}=u^{(0)\nu}$ appearing here are the unperturbed zeroth order spin and 4-velocity of the black hole. The above amplitudes, although written as an exponential are only verified by this method to third order in spin, as shown by the $+O(S^4)$ in the above expressions.

It is worth noting that there is mixing of polarization (i.e., non zero $\mathcal{M}_{+-}$ and $\mathcal{M}_{-+}$) even in absence of spin ($S^{\mu\nu}\rightarrow 0$). This is a known result and does not happen for electromagnetic wave scattering off black holes (see Ref.~\cite{Guadagnini:2002xx}).  The above amplitudes are also consistent with the scattering cross sections in the spinless case given in Ref.~\cite{Peters:1976jx}. They are also consistent with Eq.~(22) in Ref.~\cite{Guadagnini:2008ha}, valid at first order in spin. Most importantly, these amplitudes are identical to the exponentials in Eqs.~(5.14) and and (5.17) in Ref.~\cite{Aoude:2020onz} in which the amplitudes of Ref.~\cite{Arkani-Hamed:2017jhn} were rewritten as an expansion in spin multipoles in spinor-helicity formalism\footnote{Note that the conventions regarding helicity states are slightly different in Refs.~\cite{Aoude:2020onz, Arkani-Hamed:2017jhn}. They have chosen the momenta of both gravitons to be incoming. In our case, this means $l^{\mu}\rightarrow- l^{\mu}$ and the helicity of the outgoing wave/graviton is flipped. As a result, their ``same-helicity'' (++) amplitude is our helicity-reversing ($+-$) amplitude, and their ``opposite helicity'' ($+-$) amplitude is our helicity-conserving (++) amplitude. Also, they have written the results in a spinor helicity formalism. }.

The expressions for amplitudes prior to substituting the incoming and outgoing polarization vectors and for generic $C_2,~C_3$ are given in the appendix. However, after substitution of the polarization vectors, the expressions for the helicity-conserving (-reversing) amplitudes, $\mathcal{M}_{++}$, $\mathcal{M}_{--}$ ($\mathcal{M}_{+-}$, $\mathcal{M}_{-+}$) for generic $C_2$ and $C_3$ upto $S^3$ can be separately simplified by choosing the appropriate basis of vectors for writing them. For helicity-conserving amplitudes ($\mathcal{M}_{++},~\mathcal{M}_{--}$), we choose the vectors 
 \begin{alignat}{3}
\label{simsame}
 k^{\mu},~ l^{\mu},~ w_{\mr{S}}^{\mu} &= \frac{1}{2 \omega \cos^2\big(\frac{\theta}{2}\big)}[\omega(k^{\mu}+l^{\mu})	- i  \epsilon^{\mu}_{~\alpha\beta\gamma}k^{\alpha}l^{\beta}u^{\gamma}],\nnm\\ a^{\mu}&=s^{\mu}/M.
\end{alignat}
For helicity-reversing amplitudes ($\mathcal{M}_{+-},~\mathcal{M}_{-+}$), we choose the vectors
\begin{alignat}{3}
\label{simopp}
 k^{\mu},~ l^{\mu},~ w_{\mr{O}}^\mu&= \frac{-1}{2 \omega \sin^2\big(\frac{\theta}{2}\big)}[\omega(k^{\mu}-l^{\mu})+i\epsilon^{\mu}_{~\alpha\beta\gamma}k^{\alpha}l^{\beta}u^{\gamma}],\nnm\\ a^{\mu}&=s^{\mu}/M.
\end{alignat} 
In terms of these vector bases, the amplitudes for generic axisymmetric, parity-preserving, stationary compact bodies to third order in spin can be written as
\begin{widetext}
\begin{alignat}{3}
\label{keyqn2}
\nonumber \mathcal{M}_{++} &=G M \omega\frac{\cos^4(\theta/2)}{\sin^2(\theta/2)}\bigg(\exp[a\cdot ( k +l-2 w_{\mr{S}})]
+\frac{C_2-1}{2}[(k-w_{\mr{S}})\cdot a]^2+[(l-w_{\mr{S}})\cdot a]^2 
\\
& \nonumber +\frac{C_2-1}{2}[(k-w_{\mr{S}})\cdot a][(l-w_{\mr{S}})\cdot a][(k+l-2w_{\mr{S}})\cdot a]
- (C_2-1)^2[(k-w_{\mr{S}})\cdot a][(l-w_{\mr{S}}	)\cdot a](w_{\mr{S}}\cdot a) 
\\
 &  + \frac{C_3-1}{6}\{[(k-w_{\mr{S}})\cdot a]^3+[(l-w_{\mr{S}})\cdot a]^3\}\bigg). \\
\mathcal{M}_{--} & = \bar{\mathcal{M}}_{++}(a^{\mu}\rightarrow -a^{\mu}), 
\end{alignat}
\begin{alignat}{3}
 \nonumber   \mathcal{M}_{+-} &= G M \omega\sin^2\bigg(\frac{\theta}{2}\bigg) \Big(\exp[a\cdot(k-l)]+\frac{(C_2-1)}{2 \cos^2(\theta/2)}([(k-l)\cdot a]^2- \sin^2(\theta/2)\{[(k-w_{\mr{O}})\cdot a]^2+[(l+w_{\mr{O}})\cdot a]^2\\&-4 (w_{\mr{O}}\cdot a)^2\})\nonumber +\frac{(C_2-1)}{2}\tan^2(\theta/2)[(k-l-6 w_O)\cdot a][(l-w_O)\cdot a][(k+w_O)\cdot a]\nonumber \\&\nonumber+ (C_2-1)^2\tan^2(\theta/2)[(w_O+k)\cdot a][(w_O-l)\cdot a](w_O\cdot a)+\frac{(C_3-1)}{6 \cos^2(\theta/2)}([(k-l)\cdot a]^3\\&-\sin^2(\theta/2)\{[(k-w_O)\cdot a]^3-[(l+w_O)\cdot a]^3+8 (w_O\cdot a)^3\})\Big), \\  \mathcal{M}_{-+} &= \bar{\mathcal{M}}_{+-}(a^{\mu}\rightarrow -a^{\mu}).
\end{alignat}
\end{widetext}
This remarkable simplification of the Compton amplitudes in the appropriate vector basis can be helpful in the subsequent derivation of two-body dynamics. Unfortunately, the simplification for helicity-reversing amplitudes is not as nice as that of helicity-conserving amplitudes, but thus far we have not been able to find a better basis for simplification of helicity-reversing amplitudes.

 Finally, it is also worth noting that the helicity flip ($+\leftrightarrow -$) at the level of amplitudes can be carried out by reversing the direction of spin and a complex conjugation even in the most general case ($C_2,~C_3\neq 1$). This can be understood as a combination of time-reversal and parity transformation, which flips the helicities and the spin but leaves the (incoming and outgoing) momenta invariant. Thus, we have
\begin{alignat}{3}
& \mathcal{M}(\epsilon_+,\zeta_+,k^{\mu},l^{\mu}, a^{\mu}) \xrightarrow[\text{Time reversal}]{\text{Parity flip}} \bar{\mathcal{M}}(\epsilon_-,\zeta_-,k^{\mu},l^{\mu}, -a^{\mu}),\nnm\\& \\& 
\mathcal{M}(\epsilon_+,\zeta_-,k^{\mu},l^{\mu}, a^{\mu}) \xrightarrow[\text{Time reversal}]{\text{Parity flip}}\bar{\mathcal{M}}(\epsilon_-,\zeta_+,k^{\mu},l^{\mu}, -a^{\mu}),\nnm\\&
\end{alignat} 
yielding the previously seen relations $\mathcal{M}_{--} = \bar{\mathcal{M}}_{++}(a^{\mu}\rightarrow -a^{\mu})$ and $\mathcal{M}_{+-} = \bar{\mathcal{M}}_{-+}(a^{\mu}\rightarrow -a^{\mu})$.

\section{Conclusion}
\label{ftrwork}

In this work, we solved for the dynamics of a spinning compact body with spin-induced multipole moments subject to an external linearized gravitational plane wave, and solved for the scattered wave produced in response, to third order in spin of the body, at linear order in $G$. The compact body was described with an effective worldline formalism, where it was treated as a spinning point particle with non-minimal couplings of higher multipole moments to spacetime curvature. The spin-induced (electric) quadrupole and (magnetic) octupole moments of the particle could be controlled by varying two parameters $C_2$ and $C_3$, respectively, normalized so that $C_2=C_3=1$ corresponds to the black hole case. We extracted the scattering amplitude for the scattering of gravitational waves off compact spinning bodies for generic $C_2$ and $C_3$, again up to order $GS^3$. 

For the special case $C_2=C_3=1$ corresponding to a black hole, we verified that our classical scattering amplitude matches the classical limit of the Compton amplitude originating from Ref.~\cite{Arkani-Hamed:2017jhn}, in particular using its form in terms of the heavy-particle EFT variables from Ref.~\cite{Aoude:2020onz}.   
Assuming that the effective worldline theory is sufficiently general, and given that its only free parameters (contributing through order $GS^3$), $C_2$ and $C_3$, are fixed by matching to the linearized Kerr solution, this provides further evidence that the Compton amplitude from Refs.~\cite{Arkani-Hamed:2017jhn,Aoude:2020onz} is indeed suitable for describing the dynamics of a black hole, at least to third order in spin.  We note that a BCFW construction of Compton amplitudes for generic $C_2$ and $C_3$ (and so forth) has been given in Appendix B of Ref.~\cite{Chen:2021qkk}; the result for the helicity conserving (``opposite helicity'') amplitude notably differs from our result (\ref{keyqn2}) by the absence of terms quadratic in $C_2$ (``$C_{\mr S^2}$'') at third order in spin.
 We have verified that our Compton amplitudes for generic $C_2$ and $C_3$ (for both helicity configurations) are in agreement with the classical limits of the Compton amplitudes derived by the authors of Ref.~\cite{Bern:2022kto} for use in their computation of the 2-to-2 scattering amplitudes, for a certain choice of their extra free parameter (specifically when their parameter $H_2$ equals 1).

An important direction for future work is to extend our results to fourth and fifth orders in spin, where the worldline theory will encounter additional Wilson coefficients multiplying couplings quadratic in the curvature.  Input from the worldline theory may help to clarify issues such as the counting and interpretation of additional free parameters at higher orders in spin, in the comparison with effective quantum theories, and may facilitate the comparison of both of those approaches with results from black hole perturbation theory.  Beyond the (fundamentally conservative) spin-induced multipole couplings of the type considered here, it will also be necessary, at sufficiently high orders in the long-wavelength scattering expansion, to include  absorptive effects (e.g., absorption of mass-energy and angular momentum down a black hole horizon), as have been treated in an effective worldline theory (e.g.) in Ref.~\cite{Goldberger:2020fot} and using black hole perturbation theory (e.g.) in Ref.~\cite{Chatziioannou:2016kem}.

\acknowledgements

We thank Rafael Aoude, Fabian Bautista, Zvi Bern, Alessandra Buonanno, Donal O'Connell, Alfredo Guevara, Kays Haddad, Andreas Helset, Yu-tin Haung, Ted Jacobson, Gustav Jakobsen, Henrik Johansson, Chris Kavanagh, Jung-Wook Kim, Dimitrios Kosmopoulos, Andr\'es Luna, Gustav Mogull, Paolo Pichini, Jan Plefka, Radu Roiban, Jan Steinhoff and Fei Teng for helpful discussions regarding this work and/or comments on earlier versions of this paper. We thank Gustav Mogull for computing and sharing with us the classical Compton amplitude computed via worldline QFT methods up to second order in spin for comparison with our results. We thank Fei Teng for simplifying and sharing the Compton amplitudes used in Ref.~\cite{Bern:2022kto} for comparison with our results.  We are grateful to the Kavli Institute for Theoretical Physics for their hospitality at the High-Precision Gravitational Waves Program where much of this work was completed.  This research was supported in part by the National Science Foundation under Grant No. NSF PHY-1748958.
\onecolumngrid
\appendix

\section{Compton amplitudes for generic polarization vectors and generic $C_2$, $C_3$}
\label{app}

In this appendix, we write down the Compton amplitudes for generic incoming and outgoing polarization vectors. However, as with the rest of the paper, we continue to work in the traceless-transverse gauge, where $h^{\mu\nu}k_{\nu}=0$ and $h^{\mu}_{~\mu}=0$. Thus, our incoming polarization vector, denoted by $\varepsilon^{\mu}$ and the outgoing polarization vector, denoted by $\zeta_b^{\mu}$ with $a,~b=\pm 2$ satisfy $\varepsilon^2=0,~\zeta^2=0$.
For spinless bodies, the Compton amplitude to linear order in $G$ prior to substitution of polarization vectors is given by
\begin{alignat}{3}
\mathcal{M}_{ab} = \frac{4 G M \omega^3 (\varepsilon_a \cdot \zeta_b)^2}{(l-k)^2}, 
\end{alignat}
where $k^{\mu}$ and $l^{\mu}$ are the incoming and outgoing momenta/wave-vectors and $\varepsilon_a^{\mu}$ and $\zeta_b^{\mu}$ are the polarization vectors for the incoming and outgoing waves, respectively. $M$ is the mass of the compact body and $\omega$ is the frequency of the incident wave in the initial rest frame of the particle ($\omega=-k\cdot u^{(0)}$).

At linear order in spin and $G$, the addition to the Compton amplitude for generic polarization vectors is given by 
\begin{alignat}{3}
\Delta \mathcal{M}_{ab}|_{S} =& \frac{i 4 G \omega^2 (q^{\mu}S_{\mu\nu}\varepsilon_a^{\mu})(q \cdot\zeta_b)(\varepsilon_a \cdot\zeta_b)}{q^2}-\frac{i 4 G \omega^2 (k^{\mu}q^{\nu}S_{\mu\nu})(\varepsilon_a\cdot \zeta_b)^2}{q^2}  - \frac{i 4 G \omega^2 (q\cdot \varepsilon_a)(\varepsilon_a \cdot \zeta_b)(q^{\mu}S_{\mu\nu}\zeta_b^{\nu})}{q^2}
\nonumber\\
& +2 i  G \omega^2 (\varepsilon_a \cdot \zeta_b)(S_{\mu\nu}\varepsilon_a^{\mu}\zeta_b^{\nu}),
\end{alignat}
which is also independent of $C_2$, $C_3$. The amplitude is thus universal to linear order in spin. 

At second order in spin, the coefficient controlling the strength of spin-induced quadrupole moment,  $C_2$ enters the amplitude. The addition to the Compton amplitude for generic polarization vectors at $S^2$ is given by
\begin{alignat}{3}
\Delta \mathcal{M}_{ab}|_{S^2}= -C_2\frac{2G \omega^3 (q^{\mu}q^{\nu} S_{\mu}^{~\rho}S_{\nu\rho})(\varepsilon_a\cdot \zeta_b)^2}{m q^2} - \frac{2 G \omega (k^{\mu}S_{\mu\nu}\varepsilon_a^{\nu})(\varepsilon_a \cdot \zeta_b)(l^{\mu}S_{\mu\nu}\zeta_b^{\nu})}{ m } + C_2\frac{2 G \omega^3 (\varepsilon_a \cdot \zeta_b)(S_{\mu\gamma}S_{\nu}^{~\gamma}\varepsilon_a^{\mu}\zeta_b^{\nu})}{ m  }
\end{alignat}

Finally, at third order in spin, the amplitude depends on both $C_2$ and $C_3$. The addition to the amplitude at $S^3$ has terms proportional to $C_2$, $C_2^2$ and $C_3$. Interestingly, there are no terms that are independent of $C_2$ and $C_3$ at this order in spin. It is useful to separate the various contributions, since the overall expression is very long. Thus, the addition to the Compton amplitude for generic polarization vectors that is linear in $C_2$ is given by
\begin{alignat}{3}
\Delta \mathcal{M}_{ab}|_{S^3,C_2} =C_2\frac{ i G \omega^2 }{ m^2} [(S_{\mu}^{~\gamma}S_{\nu\gamma}\varepsilon_a^{\mu}\varepsilon_a^{\nu})(k\cdot \zeta_b)(l^{\mu}S_{\mu\nu}\zeta_b^{\nu})-2(k^{\mu}S_{\mu}^{~\gamma}S_{\nu\gamma}\varepsilon_a^{\nu})(\varepsilon_a \cdot \zeta_b)(l^{\mu}S_{\mu\nu}\zeta_b^{\nu})- (k^{\mu}\leftrightarrow l^{\mu}, \varepsilon_a^{\mu}\leftrightarrow \zeta_b^{\mu})].
\end{alignat}
Then, the addition to the Compton amplitude $\propto~C_2^2$ is given by 
\begin{alignat}{3}
\Delta \mathcal{M}_{ab}|_{S^3,C_2^2} = -C_2^2\frac{2i G \omega^4}{m^2}[ (S_{\mu\nu}\varepsilon_a^{\mu}\zeta_b^{\nu})(S_{\mu}^{~\gamma}S_{\nu\gamma}\varepsilon_a^{\mu}\zeta_b^{\nu}) - (\varepsilon_a\cdot \zeta_b)(S_{\mu}^{~\gamma}S_{\nu}^{~\delta}S_{\gamma\delta}\varepsilon^{\mu}_a \zeta_b^{\nu})].
\end{alignat}
Finally, the addition to the Compton amplitude at third order in spin, $\propto~C_3$ is given by
\begin{alignat}{3}
\nonumber
&\Delta \mathcal{M}_{ab}|_{S^3,C_3} = C_3\frac{i G \omega^2 }{3 m^2 q^2}[-2(q^{\mu}q^{\nu}S_{\mu}^{~\gamma}S_{\nu\gamma})(q^{\mu}S_{\mu\nu}\varepsilon^{\nu})(q\cdot \zeta)(\varepsilon \cdot \zeta)+(k^{\mu}q^{\nu}S_{\mu\nu})(q^{\mu}q^{\nu}S_{\mu}^{~\gamma}S_{\nu\gamma})(\varepsilon\cdot \zeta)^2 \\&\nonumber -q^2 (S_{\mu}^{~\alpha}S_{\nu\alpha}\varepsilon^{\mu}\varepsilon^{\nu})(k\cdot \zeta)(l^{\alpha}S_{\alpha\beta}\zeta^{\beta})+2q^2(k^{\mu}S_{\mu}^{~\gamma}S_{\nu\gamma}\varepsilon^{\nu})(\varepsilon \cdot \zeta)(l^{\alpha}S_{\alpha\beta}\zeta^{\beta}) -2q^2(l^{\mu}S_{\mu}^{~\gamma}S_{\nu\gamma}\varepsilon^{\gamma})(\varepsilon \cdot \zeta)(l^{\alpha}S_{\alpha\beta}\zeta^{\beta})\\&\nonumber -q^2(k^{\mu}k^{\nu}S_{\mu}^{~\gamma}S_{\nu\gamma})(\varepsilon\cdot\zeta)(S_{\alpha\beta}\varepsilon^{\alpha}\zeta^{\beta})+ q^2(k^{\mu}l^{\nu}S_{\mu}^{~\gamma}S_{\nu\gamma})(\varepsilon \cdot \zeta)(S_{\alpha\beta}\varepsilon^{\alpha}\zeta^{\beta}) -2q^2(l^{\mu}S_{\mu\nu}\varepsilon^{\nu})(k\cdot \zeta)(S_{\mu}^{~\gamma}S_{\nu\gamma}\varepsilon^{\mu}\zeta^{\nu})\nonumber\\&-q^2(k^{\mu}l^{\nu}S_{\mu\nu})(\varepsilon\cdot\zeta)(S_{\mu}^{~\gamma}S_{\nu\gamma}\varepsilon^{\mu}\zeta^{\nu})-q^2(k\cdot l)(S_{\mu\nu}\varepsilon^{\mu}\zeta^{\nu})(S_{\mu}^{~\gamma}S_{\nu\gamma}\varepsilon^{\mu}\zeta^{\nu})-(k^{\mu}\leftrightarrow l^{\mu},\varepsilon_a^{\mu} \leftrightarrow \zeta_b^{\mu})],
\end{alignat}
where we have written simply $\varepsilon=\varepsilon_a$ and $\zeta=\zeta_b$ for brevity.
With this, we have written down the complete expression for the Compton amplitude for generic polarization vectors for generic $C_2,~C_3$ to third order in spin. 

\twocolumngrid

%

\end{document}